\documentclass{aa}%
\usepackage{graphicx}%
\usepackage{natbib}
\bibpunct{(}{)}{;}{a}{}{,}%
\usepackage[center]{subfigure}
\begin{document}%
\title{XMM-Newton observation of the young open cluster
Blanco\,1.\thanks%
{Based on observations obtained with XMM-Newton, an ESA science mission
with instruments and contributions directly funded by ESA Member States
and NASA.}}

   \subtitle{I. X-ray spectroscopy and photometry.}

   \author{I. Pillitteri\inst{1}
          \and
          G. Micela\inst{2}
          \and
          S. Sciortino\inst{2}
          \and
          F. Damiani\inst{2}
          \and
          F. R. Harnden Jr.\inst{3}
          }

   \offprints{I. Pillitteri}

   \institute{Dip. di Astronomia, Universit\`a di Palermo,
              Piazza del Parlamento 1, 90134 Palermo - ITALY -\\
              \email{pilli@astropa.unipa.it}
         \and
             INAF - Osservatorio Astronomico di Palermo
              Piazza del Parlamento 1, 90134 Palermo - ITALY -\\
\email{giusi@astropa.unipa.it,sciorti@astropa.unipa.it,damiani@astropa.unipa.it}
         \and
          Harvard-Smithsonian Center for Astrophysics, 
          60 Garden Street, Cambridge, MA 02138\\
          \email{frh@cfa.harvard.edu}
             }

   \date{Received ; accepted }

   \abstract{We present an X-ray study  
   of the high metallicity young open cluster Blanco\,1 
   based on {\em XMM--Newton} data. 
   X-ray spectroscopy of cluster members  
   is presented for the first time as well as new 
   X-ray distribution functions of late-type stars. 
   We detected all known dF and dG stars in the EPIC field and 80\% and
90\%
   of dK and dM stars, respectively. 
   The X-ray spectral analysis of the X-ray brightest cluster stars and
X-ray color 
   analysis  of a larger sample show that a model with two temperatures 
   (at about 0.3 and 1 keV) explains the quiescent 
   activity phase spectra.
 
   We discuss also the nature of unidentified X-ray sources in the
observed region
   and their X-ray spectral properties. 
   }
\titlerunning{X-ray spectroscopy of Blanco\,1 with XMM-Newton.}

   \maketitle
\section{Introduction.}%
Open clusters are powerful laboratories to test the 
models of star formation and evolution as well as the metal 
enrichment in the Galaxy.
In fact, they are naturally selected samples of stars with same age, 
composition and environmental formation conditions.  
In the last two decades, X-ray 
observations of open clusters have assumed great importance.
Since the early '80s, the open cluster X-ray observations of first {\em
Einstein} 
and subsequently ROSAT, showed that young cluster stars are stronger 
X-ray sources than the Sun (\citealp[see][]{Micela2002} and reference
therein cited).
The evolution of X-ray activity is correlated with the evolutionary
angular momentum losses, older stars being less luminous than the
younger ones, 
but other, more subtle, factors like pre-main-sequence history and
chemical 
composition could also play a role. The comparison
of the properties of different clusters  is essential for an
understanding of the 
importance of these factors. 
It is expected that stellar chemical composition influences the 
processes of coronal emission and activity:
X-ray spectra of stars with high metal 
content are expected to have enhanced line emission contribution
with respect to low metallicity stars. Moreover,
the extent of the convective zone in solar type stars is also determined
by metal content, thus resulting in a possible enhanced dynamo
efficiency. 
Investigations of open clusters with different metallicity are of great
importance for the study of the relation between stellar structure and
X-ray activity.

Blanco\,1 is a young open cluster noticeably more distant from the 
Galactic Plane (about 240 pc) than the young open cluster scale  height
($\sim 100$ pc). 
Its age, around 100 Myr, is very similar to that of the Pleiades and
NGC\,2516
clusters \citep{deEpst85,West88}, while its metallicity ([Fe/H]=+0.23
dex)
is significantly higher than that of the Sun and the Pleiades
\citep{Edv95,Pan97,Jef99}. 
The high metallicity of Blanco\,1 offers a suitable benchmark for the
connection between stellar structure, activity and convection.
Measurements of Li abundance by \citet{Jef99} do not fit
standard mixing models for stars of this age and metallicity, 
thus implying some Li-depletion inhibition.
Our previous work (\citealp{Giusi99}; \citealp{bl1hri}, Paper I) based on 
ROSAT data, showed that the cluster dG and dK stars have an overall
X-ray emission 
similar to that of the Pleiades while the dM stars in Blanco\,1 appeared
more 
luminous in the X-ray band with respect to the Pleiades and
$\alpha$\,Per clusters.
Prior to Paper I, membership catalogues for stars down to late K types 
were based only on photometry 
\citep{deEpst85,Pan97}, while a few radial velocity measurements are
given
in \citet{Jef99} and \citet{Edv95}.
In Paper I the membership of the cluster stars in the ROSAT field of
view was 
defined by means of proper motion analysis and it was possible to 
derive X-ray luminosity distributions (XLDs) for dF, dG, dK and dM
types. 
In that work, due to the limited sensitivity of ROSAT,
the XLD of dM stars was significantly affected by the large number of
upper limits.
Furthermore, the lack of spectral capability of the ROSAT-HRI camera,
did not allow us to derive any spectral features of the cluster
coronae. The {\em XMM-Newton} observation addresses these two
issues: the large effective area of its three X-ray telescopes allows us 
to detect essentially all cluster members in the observed region
and the moderate spectral resolution of the EPIC camera allows us to
explore
the main characteristics of the coronal spectra.

In this work we present the X-ray spectroscopy and photometric analysis
of the 
cluster region as follows:
in Sect. 2 we describe the observations, the basic processing
of the data and the source detection results. 
In Sect. 3 we discuss the X-ray spectral analysis of the cluster 
stars and their XLDs. 
In Sect. 4 we discuss the nature of the unidentified sources and 
in Sect. 5 we summarize our conclusions.
\section{The observation and data analysis.}%
The observation studied here has been performed with the {\em
XMM-Newton} 
satellite in June 
2002 during revolution number 0461 with a nominal duration of 50 ksec.
The pointing, R.A.: $0^\mathrm{h}02^\mathrm{m}48^\mathrm{s}$; Dec.:
$-30^\mathrm{d}00^\mathrm{m}00^\mathrm{s}$ (J2000), 
was chosen to be the same as field 1 observed with the ROSAT HRI
\citep{Giusi99}. Because the $\sim 30^\prime$ field of view is
slightly less than that of the ROSAT HRI ($\sim 40^\prime$), the 
optical catalog of Paper I is suitable for use here.

The {\em XMM-Newton} satellite is composed of three X-ray telescopes
which
observe simultaneously, accumulating photons in three CCD-based 
instruments: the twin
MOS 1 and MOS 2 and the {\em p--n} \citep{Turner2001,Struder2001} which 
constitute the EPIC camera.
Both of these imaging and non-dispersive instruments permit low
resolution spectral analyses to be performed.
Because a fraction of the flux collected with the two telescopes in
front of the MOS CCDs is used to illuminate the dispersing grating array
RGS, the
MOS 1 and 2 collect fewer photons than does the {\em p--n}.
The {\em Medium} filter was used for 
all the three instruments during this observation.
This filter was chosen to optimize the detector response in the soft
band where it is the bulk of the coronal X-ray spectrum,
and, at the same time, to block optical photons 
(that could induce false X-ray events)
of sources as bright as 8--10 m$_V$, which are expected to be observed 
in the field of view.

\begin{figure}
\resizebox{\hsize}{!}{\includegraphics{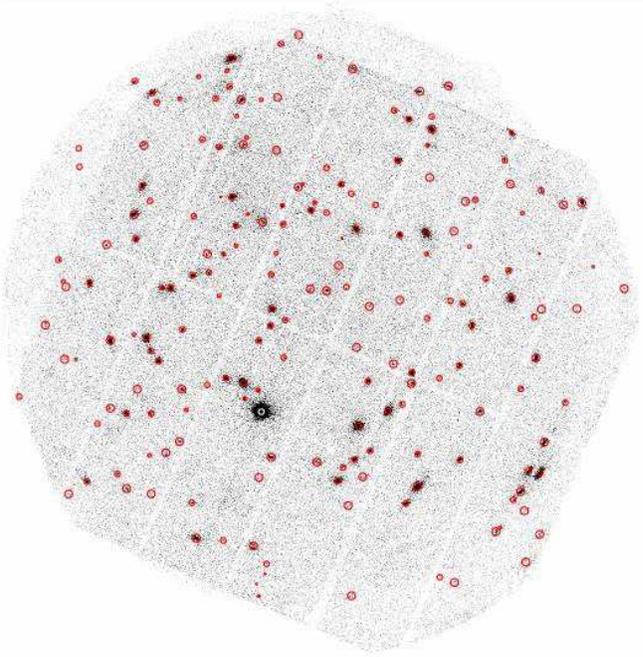}}
\caption{\label{image} Composite {\em p--n}, Mos\,1 and Mos\,2 X-ray
image of the EPIC field;
detected sources are indicated as circles with radii proportional to the
wavelet detection 
scale.}
\end{figure}
The preliminary processing of raw EPIC 
{\em Observation Data Files} data has been done using the {\em epchain} 
and {\em emchain} tasks 
of the {\em SAS} version 5.3.3 software\footnote{See http:
//xmm.vilspa.esa.es/sas}, 
which allow to calibrate both in energy and astrometry the events 
registered in each CCD chip and, finally, 
to combine them in a single data file for MOS and {\em pn} detectors.
We selected only events in the 0.3 -- 5.0 keV band, with \verb+PATTERN+
$\leq 12$ that marks only the events registered in up to four nearby 
pixels simultaneously. 
The energy upper bound reduces background contamination and
improves source detection effectiveness.
The background contribution is particularly relevant at high energies
where coronal sources have very little flux and are often
undetectable.
Due to increased solar activity, however, XMM data are sometimes
affected by an intense background level.
This behaviour degrades the quality of the data, especially for
investigations like ours that seek faint source detections,
necessitating the use of time filtering.
We have therefore selected time intervals by evaluation of a merit curve
which 
maximizes the signal-to-noise ratio and retains only time intervals with 
low image count rate (cf. \citealp{Sciortino2001}). 
Only two narrow temporal segments, comprising $\sim 750$ sec, were
affected by high count
rate, leaving most of the observation time intact for analysis
(screened fractions of total time were 48.6/49.4 and 51.0/51.7, for the
{\em p--n}
and the MOS 1 and 2, respectively).
An exposure map for each EPIC instrument was created with the {\em SAS}
task 
{\em eexpmap} in the same energy band as the selected events.
The source detection has been performed on the sum of MOS and {\em p--n} 
datasets as described in the following, by means of the Wavelet
Detection algorithm 
developed at the Osservatorio Astronomico of Palermo
(\citealp{Dami97.1,Dami97.2},
Damiani et al., in preparation) adapted to the XMM--Newton EPIC case.
The detection code has been modified in order to match the properties of 
the EPIC camera, taking into account the shape of the {\em Point Spread 
Function} (PSF) whose wings
give a relevant contribution to the encircled energy up to radii greater
than 
3 times the PSF core. The adopted PSF model follows the
parameterization described in \citet{GhizzardiPSF} and \citet{Saxon02},
valid out to $\sim 12 \arcmin$ off-axis. 
The code avoids the problem of spurious peaks in the convolved image due
to sharp CCD edges
by interpolating
the image prior of convolution in the gaps and extrapolating beyond the 
boundaries with the nearest-to-local background.
Following detection, count rates are obtained multiplying by 
the area of a reference instrument chosen by the user, 
as if the observation were carried with that single instrument. 
We chose the MOS1 and calculated our count rate to flux 
conversion factor for the {\em Medium}
filter used during our observation\footnote{In these data accumulated 
{\em p--n} events are $\sim$3.2 times those recorded for each MOS
detector}.
\begin{table*}
\renewcommand{\arraystretch}{1.1}
\renewcommand{\tabcolsep}{0.2cm}
\small
\begin{center}
\caption{\label{xmemb}
X-ray properties of Blanco\,1 stars in the {\em XMM-Newton} -- EPIC
field. 
Identifiers from \citet{Pan97} (ZS), from \citet{Giusi99} (BLX). We
named here P stars those without identifier from Paper I.
Positions and B--V color are taken from Paper I; 
exposure times and count rates are scaled on the MOS1 instrument;
fluxes are calculated with a conversion factor of  $5.69\ 10^{-12}$ ergs
cts$^{-1}$ cm$^{-2}$ (see Sect. 3.2).}
\begin{tabular}{l l l l l l l l l l l}\hline \hline
Name& RA & Dec & Offset & Signif &B--V &Exp. Time & Count rate & $\log$
flux & $\log \mathrm{L_X}$ & $\log
\frac{\mathrm{L}_\mathrm{X}}{\mathrm{L}_\mathrm{bol}} $ \\ 
  &  J2000 & J2000 & arcsec &  &  & ksec. & cts. ksec.$^{-1}$ & ergs
s$^{-1}$cm$^{-2}$& ergs s$^{-1}$ & \\\hline
P1   & 0:01:37.8 & -29:57:28.3 & --  &  --  & 1.58 & 26.28 & $\leq$1.9
&  $\leq$-14.15  &  $\leq$28.72 &$\leq -2.77$ \\
P2   & 0:01:52.6 & -30:05:36.0 & --  &  --  & 1.42 & 49.46 & $\leq$0.9
&  $\leq$-14.27  &  $\leq$28.60 &$\leq -3.61$ \\
ZS37  & 0:01:53.4 & -30:06:12.9 &  0.66 & 23.4 & 1.48 & 47.05 & 8.0
$\pm$ 0.6 & -13.38 & 29.50 & -2.49\\
ZS38  & 0:01:54.4 & -30:07:42.0 &  2.11 & 39.25 & 1.04 & 44.98 & 16.9
$\pm$ 0.9 & -13.06 & 29.82 & -3.09\\
P3   & 0:01:54.5 & -30:10:38.3 &  2.06 & 10.52 & 2.02 & 25.41 & 4.3
$\pm$ 0.7 & -13.65 & 29.22 & --\\
ZS40  & 0:01:56.9 & -30:12:08.0 &  11.57 & 7.22 & 1.38 & 25.19 & 2.9
$\pm$ 0.5 & -13.83 & 29.05 & -3.31\\
ZS39  & 0:01:57.7 & -30:09:28.8 &  2.11 & 7.84 & 0.1 & 30.11 & 2.3 $\pm$
0.5 & -13.92 & 28.96 & -5.81\\
BLX7  & 0:02:00.8 & -29:59:17.6 &  0.95 & 40.32 & 0.74 & 69.05 & 11.4
$\pm$ 0.6 & -13.23 & 29.65 & -3.79\\
BLX9  & 0:02:01.3 & -29:57:55.3 &  1.14 & 13.57 & 1.5 & 67.89 & 2.0
$\pm$ 0.2 & -13.97 & 28.90 & -3.01\\
ZS43  & 0:02:03.7 & -30:10:25.1 &  9.03 & 5.12 & 1.53 & 33.99 & 0.6
$\pm$ 0.2 & -14.52 & 28.35 & -3.44\\
ZS42  & 0:02:04.2 & -30:10:34.5 &  3.64 & 19.66 & 1.35 & 32.62 & 6.9
$\pm$ 0.7 & -13.44 & 29.43 & -2.98\\
ZS45  & 0:02:18.5 & -29:51:08.6 &  0.8 & 34.55 & 0.66 & 66.08 & 8.9
$\pm$ 0.5 & -13.33 & 29.54 & -4.10\\
ZS46  & 0:02:19.7 & -29:56:07.6 &  0.66 & 64.35 & 1.17 & 94.38 & 14.3
$\pm$ 0.5 & -13.13 & 29.75 & -3.00\\ 
ZS48  & 0:02:21.6 & -30:08:21.7 &  0.72 & 56.5 & 0.27 & 59.65 & 22.2
$\pm$ 0.9 & -12.94 & 29.94 & -4.50\\
P4   & 0:02:24.3 & -30:06:02.9 &  -- &  --  & 1.40 & 84.87 & $\leq$0.4
&  $\leq$-14.58  &  $\leq$28.29 & $\leq -3.98$\\
ZS53  & 0:02:24.3 & -30:09:09.0 &  1.38 & 27.95 & 1.53 & 63.4 & 5.1
$\pm$ 0.4 & -13.57 & 29.30 & -2.51\\
BLX17 & 0:02:25.9 & -29:52:39.2 &  2.05 & 39.11 & 1.42 & 82.75 & 7.2
$\pm$ 0.4 & -13.43 & 29.45 & -2.77\\
ZS54  & 0:02:28.2 & -30:04:43.6 &  0.22 & 57.26 & 0.94 & 97.82 & 11.9
$\pm$ 0.5 & -13.21 & 29.67 & -3.39\\
ZS61  & 0:02:34.8 & -30:05:25.6 &  0.83 & 79.74 & 0.84 & 102.3 & 18.2
$\pm$ 0.6 & -13.02 & 29.85 & -3.38\\
ZS62  & 0:02:35.4 & -30:07:02.0 &  2.13 & 31.12 & 0.63 & 71.92 & 6.1
$\pm$ 0.4 & -13.5 & 29.38 & -4.34\\ 
BLX24 & 0:02:48.4 & -29:53:53.8 &  4.33 & 23.53 & 1.58 & 75.28 & 4.0
$\pm$ 0.4 & -13.68 & 29.20 & -2.31\\
BLX26 & 0:02:51.5 & -29:54:49.4 &  2.04 & 19.11 & 1.52 & 105.96 & 1.6
$\pm$ 0.2 & -14.07 & 28.8 & -3.04\\
P5   & 0:02:52.2 & -29:47:00.9 &  3.2 & 10.66 & 1.47 & 44.29 & 2.1.0
$\pm$ 0.3 & -13.95 & 28.92 & -3.10\\
BLX27 & 0:02:54.2 & -30:06:55.9 &  0.7 & 23 & 1.42 & 95.04 & 3.6 $\pm$
0.3 & -13.72 & 29.15 & -3.07\\
ZS77$^a$& 0:02:55.1 & -30:08:53.8 & -- & -- & 0.03 & 58.1 & $\leq0.6$ &
$\leq-14.44$ & $ \leq 28.43$ & $\leq -6.55$\\
ZS76$^b$& 0:02:56.4 & -30:04:45.1 &  0.24 & 226.64 & 0.76 & 113.81 &
110.3 $\pm$ 1.3 & -12.24 (-12.68) & 30.63 (30.19) & -2.76 (-3.20)\\
ZS75  & 0:03:00.3 & -30:03:21.8 &  0.26 & 65.99 & 0.79 & 121.93 & 12.0
$\pm$ 0.4 & -13.21 & 29.67 & -3.67\\
ZS71  & 0:03:02.9 & -29:47:44.2 &  0.25 & 47.78 & 1.37 & 58.22 & 20.4
$\pm$ 0.8 & -12.97 & 29.90 & -2.48\\
ZS84  & 0:03:10.8 & -30:10:49.1 &  1.64 & 35.03 & 0.52 & 58.35 & 11.8
$\pm$ 0.6 & -13.21 & 29.66 & -4.32\\
BLX37 & 0:03:11.5 & -29:58:10.2 &  1.33 & 22.75 & 1.51 & 102.68 & 2.7
$\pm$ 0.2 & -13.85 & 29.02 & -2.87\\
ZS95  & 0:03:16.5 & -29:58:47.7 &  1.46 & 33.16 & 0.75 & 109.05 & 4.6
$\pm$ 0.3 & -13.62 & 29.25 & -4.17\\
ZS91  & 0:03:20.6 & -29:49:22.9 &  0.92 & 25.72 & 0.44 & 57.84 & 6.5
$\pm$ 0.5 & -13.47 & 29.40 & -4.73\\
P6   & 0:03:20.8 & -29:51:52.8 & --  &  --  & 1.36 & 71.62 & $\leq$0.6
&  $\leq$-14.43  &  $\leq$28.45 & $\leq -3.95$\\
ZS96  & 0:03:21.8 & -30:01:10.8 &  0.54 & 45 & 0.25 & 95.73 & 8.2 $\pm$
0.4 & -13.37 & 29.50 & -4.99\\
P7   & 0:03:22.5 & -29:51:52.8 &  2.57 & 11.52 & 1.45 & 69.89 & 2.0
$\pm$ 0.3 & -13.98 & 28.9 & -3.21\\
ZS94  & 0:03:24.2 & -29:56:23.1 &  2.22 & 22.3 & 1.39 & 79.39 & 3.5
$\pm$ 0.3 & -13.74 & 29.13 & -3.19\\
ZS90  & 0:03:24.4 & -29:48:49.6 &  0.72 & 20.09 & 0.33 & 11.56 & 14.0
$\pm$ 1.7 & -13.14 & 29.74 & -4.61\\
ZS93  & 0:03:24.7 & -29:55:14.8 &  0.75 & 32.3 & 0.93 & 79.72 & 7.1
$\pm$ 0.4 & -13.43 & 29.44 & -3.63\\
BLX46 & 0:03:34.5 & -29:58:30.5 &  0.9 & 26.42 & 1.56 & 75.84 & 4.4
$\pm$ 0.3 & -13.64 & 29.24 & -2.28\\
P8   & 0:03:39.9 & -29:58:45.0 &  2.34 & 5.45 & 1.45 & 69.21 & 0.8 $\pm$
0.2 & -14.38 & 28.49 & -3.62\\
ZS107 & 0:03:50.2 & -30:03:55.9 &  4.45 & 7.46 & 1.15 & 21.55 & 2.4
$\pm$ 0.5 & -13.9 & 28.97 & -3.80\\ \hline
\end{tabular}\\
\end{center}
$^a$: star added from \citet{Pan97} catalog, see text.\\
$^b$: in parentheses flux, $\log \mathrm{L_X}$  and $\log
\mathrm{L_X}/\mathrm{L}_\mathrm{Lbol}$ of quiescent phase.
\end{table*}
In order to choose detection thresholds, we performed a set of 
100 simulations of empty fields with the same number of counts as in our
observation. 
We then chose the threshold to retain at most only a single
spurious source.
After removal of
several obvious spurious detections\footnote{These were due to ``hot
pixels'' not yet listed in calibration files at the time
our work has been done.}, 190 detected sources remained.

An optical catalog of 93
Blanco\,1 members was derived in Paper I from GSC-II
\footnote{The Guide Star Catalogue II is
a joint project of the Space Telescope Science Institute and the INAF - 
Osservatorio Astronomico di Torino.}  data by means of
proper motion and 
photometric selection in order to avoid any bias toward X-ray properties 
of the sample. 
That analysis was also restricted to the field of view of 
ROSAT-HRI observation in order to maximize the effectiveness of
selection. 
Forty of these 93 objects fell in the EPIC field of view.
To this catalog we added the bright A-type star that lies in the EPIC
field of view,
assuming this star to be a member \citep{Pan97}, notwithstanding its
lack of proper-motion
based membership.

While cross-identifying X-ray source and optical catalog positions, we
found
systematic offsets of $2.2\arcsec$ and $0.4\arcsec$ along R.A. and
Dec., respectively, requiring that corrections be applied to the X-ray
positions
before definitive matches could be achieved.
In this second iteration, we retained identifications for which the 
offset between X-ray and optical positions was $\leq 13\arcsec$.
This cross-identification radius represents a reasonable choice for
minimizing the number of spurious identifications 
(whose expected number at $13\arcsec$ is 0.86)
while achieving the largest number of bona-fide identifications.
We thus find 36 X-ray sources with optical cluster member 
counterparts and report their properties in Table \ref{xmemb}.
Exposure times and count rates listed in Table \ref{xmemb} are scaled 
to the reference instrument (MOS\,1) that we chose for detection. 
For the five cluster stars that were not detected, we have calculated
upper limits
to count rate, flux and luminosity, using the chosen detection
threshold.

A search for optical counterparts of the remaining 154 X-ray sources,
using the USNO-B1 \citep{Monet03}, GSC-II
and 2MASS%
\footnote{2MASS is a joint project of the University of Massachusetts
and the 
Infrared Processing and Analysis Center/California Institute of
Technology, 
funded by the National Aeronautics and Space Administration and the
National 
Science Foundation.} catalogs \citep{Cutri2003} with a match
radius of 13$\arcsec$ yielded a list of 90 counterparts in the visible
and infrared bands. We report the X-ray properties of these 90 sources
in Table A.1,
in which the last 3 columns show the number of matches in the three
catalogs. 
Note that the USNO catalog
yields an implausible number of matches for some sources, indicating
problems with this catalog. 
In such cases, an extended object (e.g., a galaxy) can often be
recognised in the finding
chart.

After the identification process, 64 of 190 sources remain without
counterparts.  
Due to the high galactic latitude of this pointing, we expect to have
detected
a small number of active field stars and
a larger number of X-ray extragalactic sources (e.g., AGN, galaxies or
galaxy clusters)
Properties of these unidentified X-ray sources will be discussed in
Sect. 4.

\section{Blanco\,1 X-ray sources.}
\subsection{X-ray Spectral Analysis.}%
The spectral capabilities of the EPIC camera allow us to explore
the coronal properties of cluster stars. 
We have used {\em p--n} data and \verb+XSPEC+ software to perform spectral 
analyses for six cluster stars (3 dK, 2dG and 1 late A)
with more than 1000 X-ray counts,
identifying
relevant coronal temperature components and corresponding 
{\em emission measures} (EM).
Spatial isolation of these sources
minimized contamination by other sources.
We then used the results of this analysis as a guideline for our
subsequent 
analysis with X-ray color indices (see Sect. 3.2).

\begin{table*}
\begin{center}
\caption[]{\label{bfit}Best fit parameters derived from 
the spectral analysis of 
six cluster brightest X-ray sources. The spectral types are estimated
from B--V
assuming Main-Sequence stars \citep[see][]{Langbook}. 
The error ranges refer to $\pm 1\ \sigma$ level.}
\renewcommand{\arraystretch}{1.3}
\begin{tabular}{l l l l l l l l l}\hline \hline
Name & Sp. Type & T$_1$ (keV) & T$_2$ (keV) & $Z/Z_{\sun}$ & $\log$
EM$_1$ (cm$^{-3}$)& $\log$ EM$_2$ (cm$^{-3}$) & $\chi_\nu^2$ (dof)
&P($\chi > \chi_o$) \\ \hline
$ZS 46$& K5 & $0.32^{+0.02}_{-0.03}$ & $1.04^{+0.04}_{-0.07}$ &
$0.23^{+0.08}_{-0.05}$ & $52.58^{+0.09}_{-0.10}$ &
$52.52^{+0.08}_{-0.08}$ & 1.16 (34) & 23.5\% \\ 
$ZS 48^a$& A8 & $0.60^{+0.02}_{-0.02}$ & --  & $0.9^{+0.1}_{-0.2}$ &
$52.61^{+0.09}_{-0.1}$ & -- & 1.48 (26) & 5.6\% \\ 
$ZS 54$& K2 & $0.40^{+0.03}_{-0.05}$ & $0.90^{+0.16}_{-0.04}$ &
$0.21^{+0.08}_{-0.05}$ & $52.54^{+0.12}_{-0.07}$ &
$52.44^{+0.14}_{-0.11}$ & 1.25 (31) & 15.7\% \\ 
$ZS 61$& K0 & $0.33^{+0.02}_{-0.04}$ & $1.06^{+0.07}_{-0.07}$ &
$0.22^{+0.06}_{-0.05}$ & $52.72^{+0.04}_{-0.09}$ &
$52.57^{+0.08}_{-0.08}$ & 0.78 (46) & 85.1\% \\ 
$ZS 75$& G8 & $0.29^{+0.03}_{-0.02}$ & $1.02^{+0.09}_{-0.08}$ &
$0.17^{+0.07}_{-0.04}$ & $52.63^{+0.09}_{-0.10}$ &
$52.52^{+0.07}_{-0.08}$ & 1.02 (34) & 43.6\% \\ 
$ZS 76^b$& G8 & $0.35^{+0.05}_{-0.03}$ & $0.99^{+0.04}_{-0.04}$ &
$0.22^{+0.05}_{-0.05}$ & $53.03^{+0.09}_{-0.32}$ &
$53.12^{+0.03}_{-0.12}$ & 0.82 (72) & 87.8\% \\ \hline
\end{tabular}\\
\end{center}
$^a$: For ZS48 the model is 1-T \verb+VAPEC+, $Z/Z_{\sun}$ refers to Fe
abundance, O abundance is $0.4^{+0.2}_{-0.02}$.\\
$^b$: For $ZS 76$ we report the quiescent spectrum parameters. 
\end{table*}
\begin{figure*}
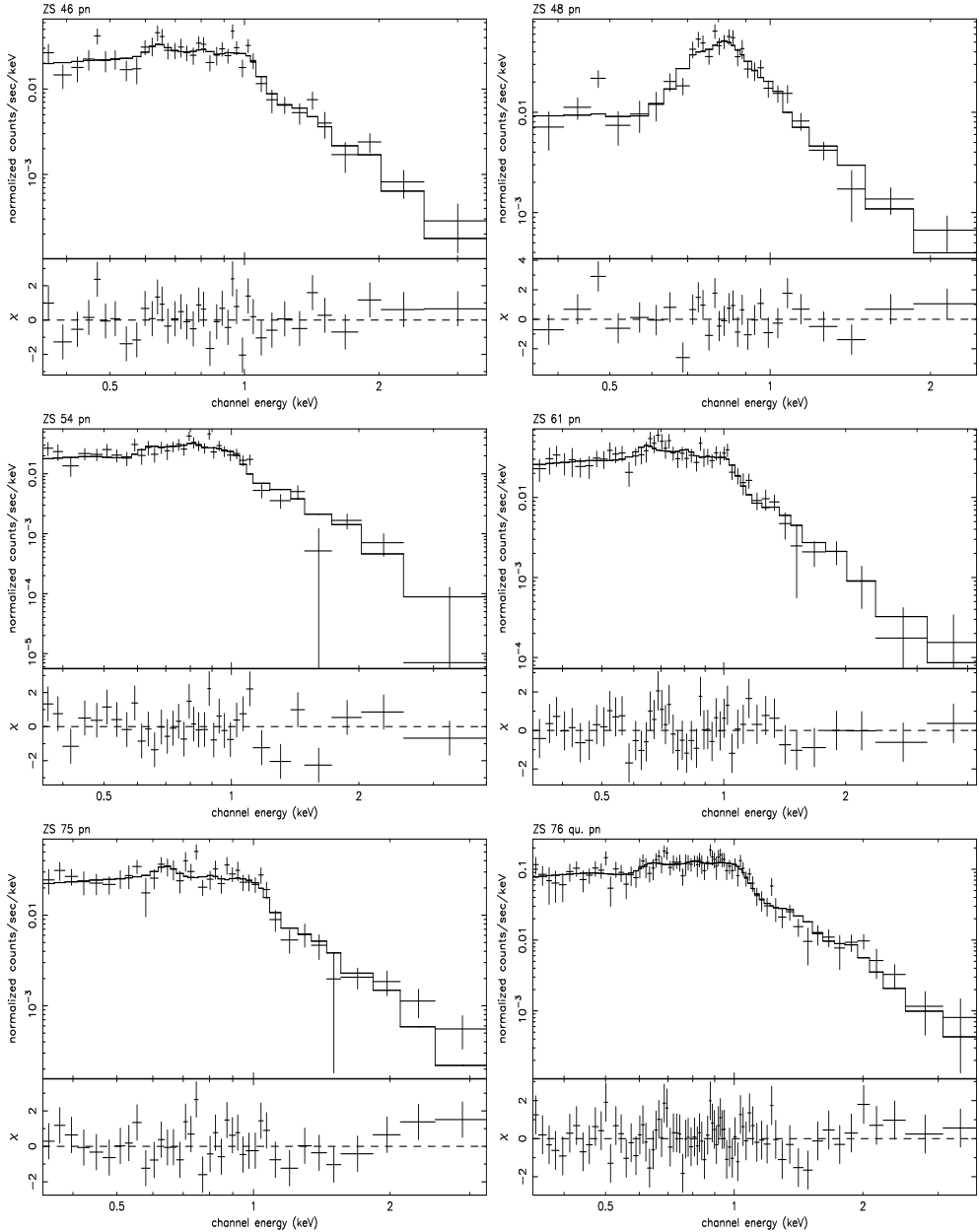

\begin{center}
{\includegraphics[height=6.5cm,width=5.5cm,angle=-90]{zs46spec.ps}}
{\includegraphics[height=6.5cm,width=5.5cm,angle=-90]{zs48spec.ps}}\\
{\includegraphics[height=6.5cm,width=5.5cm,angle=-90]{zs54spec.ps}}
{\includegraphics[height=6.5cm,width=5.5cm,angle=-90]{zs61spec.ps}}\\
{\includegraphics[height=6.5cm,width=5.5cm,angle=-90]{zs75spec.ps}}
{\includegraphics[height=6.5cm,width=5.5cm,angle=-90]{zs76spec-qu.ps}}\\
\end{center}
\caption{\label{spec} X-ray spectra of the Blanco\,1 brightest X-ray
sources.} 
\end{figure*}
The star {\em ZS\,76} underwent two flare-like episodes 
during the first half of our observation. This star is reported as 
binary in the catalog of \citet{Pan97}.
Although the beginning of the first flare may not have been observed,
the two flares appear to have durations of $\sim 4$ hours.
Here we present our analysis of the quiescent phase;
the flare spectra will be analyzed in detail in a companion paper
devoted to a study of the X-ray variability of the cluster.
Stars {\em ZS46}, {\em ZS61} and {\em ZS75} exhibit no flares, although
Kolmogorov-Smirnov tests indicate some type
of variability at the 98\%-confidence level;
we will discuss these features  in the context of our variability study.
{\em ZS48} and {\em ZS54} do not show variability.
Here we assume that the spectra of these five stars are 
representative of their normal, quiescent X-ray activity. 

For our spectral analyses,  we extracted events that triggered only one
or two detector pixels at the same time
(\verb+PATTERN+ $\leq 4$),
with energy in 0.3--5.0 keV, band and in a circular region 
within 35 or 40$\arcsec$ from the position of the sources depending on
the spatial crowding. Spectra were binned to provide at least 25 counts
per bin.
In some cases for the background spectrum we chose an extraction region
near but not centered on the source to avoid 
contamination from other X-ray sources. 
Redistribution matrix and response files were calculated at each source
position.
We fitted the data with a combination of two absorbed plasma thin
emission models; a combination of 3 components did not improve the fits. 
To calculate the models and to fit the data, we used the \verb+APEC+ database,
which contains the relevant atomic physics data for both continuum and line 
emission \citep{Smith2001}, driven by \verb+XSPEC+ software.

In all cases we found $N_\mathrm{H}$ $\sim 3\ 10^{20}$ cm$^{-2}$
to be consistent with A$_V$ from photometry estimates 
\citep{Epst68,deEpst85,West88} and therefore fixed it at this value. 
Abundances  were scaled on the solar values of \citet{Angrev89}, leaving
as
free parameters the temperatures, a global abundance value and the
emission measures (EM) of the two thermal components.
Our best-fit parameters are reported in Table \ref{bfit} and the
corresponding spectra
are shown in Fig. \ref{spec}. 
For late type stars (i.e., $ZS46, ZS54, ZS61, ZS75, ZS76$)
the 2-T model describes the shape of
coronal spectrum quite well and indicates main temperature components
around 0.3 and 1.0 keV. The ratio of emission measures $EM_2/EM_1$
is in the 0.7--1.2 range. 
The global coronal abundance parameter, determined mostly by the Fe line
complex
around 1 keV, is always significantly lower than the solar value.
Similar results for temperature value, emission measure ratio and
coronal abundances 
are found by \citet{Briggs2003} in an analogous spectral 
analysis of Pleiades solar type stars observed with XMM-Newton.
\citet{Gagne95} previously found a 2-T model to be required to explain 
ROSAT PSPC spectra of Pleiades solar type stars, 
although emission measure ratios
found in that work were greater than 1, possibly indicating a more
significant hot component.
The need for lower than solar coronal 
abundances in the Blanco\,1 coronae could imply that photospheric
composition
doesn't constrain spectral features of the coronal plasmas, but
fitted emission measures are entangled with the metallicity values
derived in this global fitting
analysis, thus preventing a definitive abundance study. 
For Pleiades stars HII 1032, 1100, 1348 and 1516, studied by Briggs \&
Pye (cf. their Table 3) with a 2-T analysis in the same spectral range,
the mean EMs of hot and cool components are on average $\sim0.3$ dex 
lower than those of our sample.
By including also HII 1110 and 1280, two dK late stars with less than 
500 X-ray counts, we found a more marked difference. 
However, we discarded these stars in the comparison because of their poor 
statistics. 
A comparison of our L$_\mathrm X$ values for Blanco\,1 with those for
the Pleiades observations indicates that different parts of the
XLDs are being sampled in the two cases: we are modeling sources in the
high activity tail 
($\log L_\mathrm{X} \sim 29.7$) while Briggs \& Pye studied stars over a
wider range of luminosity. This likely explains the differences noted above.

The spectrum of the A8 star $ZS48$ differs
from that of solar type stars, and is well fit
with a 1-T model with kT $\sim$ 0.6 keV\footnote{Note however that the
EPIC {\em p--n} bandpass is rather insensitive to any cool component
below 0.3 keV.}.
We allowed
abundances of Fe and O to vary (see caption of Table \ref{bfit})
and kept all other metallicities fixed at the solar value. 
Stellar structure models predict that an A8 star with Blanco\,1
metallicity
would have a very shallow convective zone \citep{Siess2000} and hence
would exhibit very little X-ray emission.
In the case of a binary system
unresolved in X rays, emission is often assigned to the late 
companion. 
Because this star is flagged as a possible binary in the Guide Star
Catalog II
\footnote{Available through the Vizier web database: 
http://vizier.u-strasbg.fr },
a late companion could be
responsible of the X-ray spectrum. If this were the case, however,
its spectrum should resemble those of other late type stars. 
We therefore suggest instead that the emission is intrinsic.
\subsection{\label{color} X-ray colors, fluxes and luminosities.}
Although the spectra of most cluster members have relatively poor 
statistics, we can characterize their coronal emission by means
of {\em X-ray colors indices} and a grid of optically-thin coronal
plasma
models; a similar analysis method was employed by \citet{Damiani2003}
but on {\em Chandra} ACIS-I data.
We chose three bands in which to accumulate
EPIC source photons {\em p--n} detector ({\em soft}: 0.3-0.75 keV;
{\em medium}: 0.75-1.5 keV and {\em hard}: 1.5-5.0 keV) and calculated
{\em colors} as follows:
\[ CR1 = 2.5 \log \frac{N_{medium}}{N_{soft}} \]
\[ CR2 = 2.5 \log \frac{N_{hard}}{N_{medium}} \]
These bands were chosen to minimize errors in each band.
Source photons were extracted from circular regions of 20$\arcsec$
radius, and
background was obtained from annuli with radii of 22.5$\arcsec$ 
and 30$\arcsec$.
Smaller extraction radii than for detection are needed here
to minimize contamination from nearby sources, especially in 
the case of the faint sources.
Differences of encircled energy at the chosen radii
between soft, medium and hard bands are negligible ($\leq 2\%$) and were 
therefore ignored. 

\begin{figure}
\resizebox{\hsize}{!}{\includegraphics{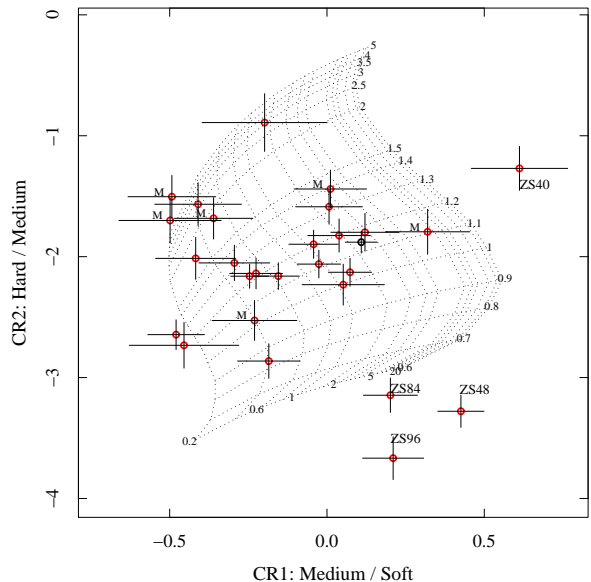}}
\caption{\label{hrfig} $CR2$ vs. $CR1$ scatter plot for the cluster
X-ray sources
based on {\em p--n} data. 
The region explored by the model grid is traced; values of EM$_2$/EM$_1$
are indicated at the bottom of
the grid, and high temperature component values (units: keV) are 
indicated at the right edge.
``M'' denotes dM type stars, while stars outside
the grid and discussed in the text are labelled with identifiers.}
\end{figure}

We computed a grid of models with \verb+XSPEC+, adopting a 
2-T \verb+APEC+ model with photoelectric absorption. 
To reduce the number of parameters we used the above X-ray spectroscopy
results as a guideline, fixing the absorption
$N_{\mathrm H}$ at $3\cdot 10^{20}\ \mathrm{cm}^{-2}$,
the lower temperature at 0.33 keV
and the abundances $Z= 0.25Z\sun$
(the median of values obtained in the detailed spectral fits);
this left high temperature, in the range 
0.6-5.0 keV, and EM ratio of the two \verb+APEC+ components (range
0.2-20) 
as the only two free parameters.
We then calculated the expected
$CR1$ and $CR2$ colors over a grid of 240 spectra.
Table \ref{hrtab} reports
the observed $CR1$ and $CR2$,
while Fig. \ref{hrfig} shows the corresponding points in the
$CR1$ -- $CR2$ plane. 
All cluster sources have $CR2$ color indices below zero,
while the $CR1$ index spans values around zero.
Most of the cluster sources are included in the region 
explored by the model grid and broad agreement is found for the
stars analyzed in detail in the Sect.3.1.
Star {\em ZS48} lies outside the grid, as expected because its spectrum
was adequately fit with a 1-T model at 0.6 keV.
Analogously, the position of {\em ZS96} in the diagram 
suggests that a 2-T model is not required to describe this A type star.
Apparently discrepant CR values of the G star {\em ZS84} are
probably due to the presence of a nearby object,
as suggested by an inspection of the optical finding chart.
Note too that the far-off axis position of {\em ZS40}
and the low statistical precision of its spectrum have probably
invalidated its CR estimates.  
In summary, a significant fraction of the sample is found
to have a hot temperature component between 0.8 and 1.5 
keV, with weights generally lower than (or at most equal to) the cool
component.
Some M type stars are also found at higher temperature 
($\sim$1.5--5 keV, see Fig. \ref{hrfig}).
The agreement of this CR method with the spectral analysis suggests that
the 
CRs offer a good description of the coronal spectra in
terms of 2-T models.
  
\begin{table}
\renewcommand{\arraystretch}{1.1}
\centering
\caption{\label{hrtab} X-ray colors $CR1$ and  $CR2$ in the bands:
0.3--0.75,
0.75--1.5, 1.5--5.0 keV defined in Sect. 3.2 for the sample inside the
grid
of Fig.\ref{hrfig}. The last column reports the count to flux conversion 
factor in units of $10^{-12}$ ergs cts$^{-1}$cm$^{-2}$.}
\begin{tabular}{l l l l}\hline \hline
Name	 & 	$CR1$	 & 	$CR2$	 & 	CF   \\ \hline
BLX24 & -0.49 & -1.5 & 5.64 \\
BLX27 & -0.41 & -1.57 & 5.81 \\
BLX9 & -0.5 & -1.7 & 5.64 \\
BLX46 & -0.23 & -2.53 & 5.32 \\
ZS53 & -0.36 & -1.68 & 5.7 \\
ZS94 & -0.42 & -2.01 & 5.49 \\
ZS93 & -0.29 & -2.05 & 5.61 \\
ZS61 & -0.25 & -2.16 & 5.57 \\
ZS71 & -0.23 & -2.14 & 5.57 \\
ZS75 & -0.15 & -2.16 & 5.57 \\
ZS54 & -0.03 & -2.06 & 5.76 \\
ZS95 & -0.48 & -2.64 & 5.14 \\
ZS62 & -0.18 & -2.86 & 5.22 \\
ZS39 & -0.2 & -0.89 & 7.12 \\
BLX37 & 0.01 & -1.44 & 6.3 \\
BLX17 & 0.01 & -1.59 & 6.22 \\
ZS76 & 0.32 & -1.79 & 6.65 \\
BLX7 & 0.04 & -1.83 & 6.13 \\
ZS38 & -0.04 & -1.9 & 6.13 \\
ZS45 & 0.12 & -1.8 & 5.82 \\
ZS37 & 0.32 & -1.79 & 6.33 \\
ZS46 & 0.07 & -2.13 & 5.69 \\
ZS42 & 0.05 & -2.23 & 5.69 \\ \hline
\end{tabular}
\end{table}
By evaluating the flux obtained from the best--fit spectral models we
derived 
a conversion factor CF between count rate and flux 
(cf. Table \ref{hrtab}).  
The median value is $5.69\ 10^{-12}$ ergs cts$^{-1}$ cm$^{-2}$, with a
1-$\sigma$ 
range of 5.52 -- 6.25 $10^{-12}$ ergs cts$^{-1}$ cm$^{-2}$ 
corresponding to uncertainties of -3\% and +10\%, respectively.
We calculated X-ray fluxes and luminosities
in the 0.3--5.0 keV band (cf. Table \ref{xmemb}) assuming a distance of 
250 pc \citep{Pan97,West88,deEpst85} as in previous works. 
Note however that the cluster distance derived from the {\em Hipparcos}
satellite 
is 262 pc \citep{Robich99}, which would lead to a difference of $\sim
+0.02$ in $\log \mathrm{L_X}$.
Table \ref{xmemb} gives count rates, fluxes and luminosities of
the cluster sample. Count rate errors calculated by the detection code
have a median value of $\sim$ 7\% (4\% and 19\% at the 0.1 and 0.9
quantiles of distribution, respectively).

\begin{figure}
\resizebox{\hsize}{!}{\includegraphics[angle=-90]{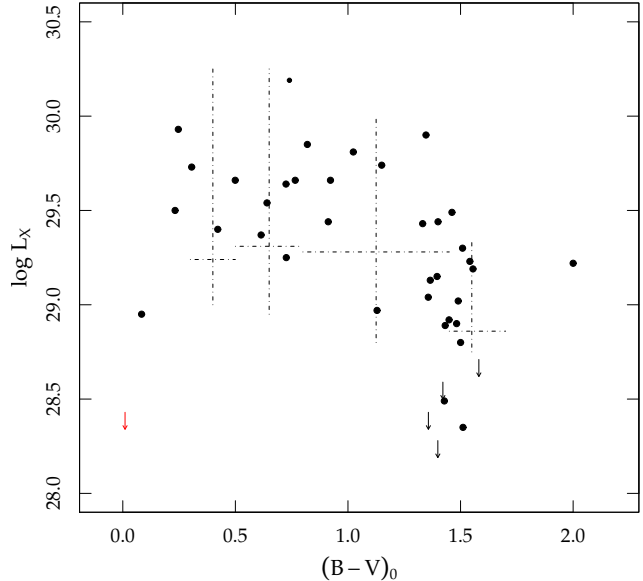}}
\caption{\label{lxbv} $\log \mathrm{L}_\mathrm{X}$ vs. $(B-V)_0$ scatter
plot 
for the Blanco\,1 X-ray sources.
Horizontal lines are medians of $\log \mathrm{L}_\mathrm{X}$ for the
Pleiades, while 
vertical lines are the 0.1--0.9 quantiles of Pleiades data in each 
spectral type.}
\end{figure}

In Fig. \ref{lxbv} we plot Blanco\,1 $\log \mathrm{L}_\mathrm{X}$ vs.
$(B-V)_0$ 
and for comparison, median and 0.1--0.9 quantiles of $\log
\mathrm{L}_\mathrm{X}$
for the Pleiades \citep{Giusi999,Giusi96}.
The new data permit us to detect 11 X-ray faint members for which only
upper limits were obtained with ROSAT, giving
an XMM detection percentage of 88\% vs. 61\% for ROSAT.

Three A-type stars of the Paper I catalog are detected, but
{\em ZS 77} (the star added from Panagi's list) is not and instead
we estimate its upper limit.
All dF and dG type stars in the EPIC field of view (numbering 3 and 6,
respectively) were detected, and the
dK and dM type detection fractions were
15/18 and 9/10 (83\% and 90\%), respectively.

Figure \ref{lxbv} suggests a rise of X-ray luminosities
at $(B-V)_0 \sim$ 0.1 possibly indicating
the onset of convection. The
dynamo mechanism responsible for coronal emission may start at these
spectral types, corresponding to stars with masses $\sim$2M$\sun$.
For comparison, ROSAT nearby star data (\citealp{Schmitt97})
show detections at similar types ($B-V \sim$ 0.2). 
X-ray luminosities have a spread of $\sim 0.5 - 0.7$ dex in dF, dG and
early dK
types. Furthermore, a drop of about 1 dex in $\log
\mathrm{L}_\mathrm{X}$ is 
visible in late dK and dM stars with the 
lowest detected source having a flux of $\log f_\mathrm{X} = -14.52$ 
corresponding to $\log \mathrm{L_X} = 28.35$. 

\begin{figure}
\resizebox{\hsize}{!}{\includegraphics[angle=-90]{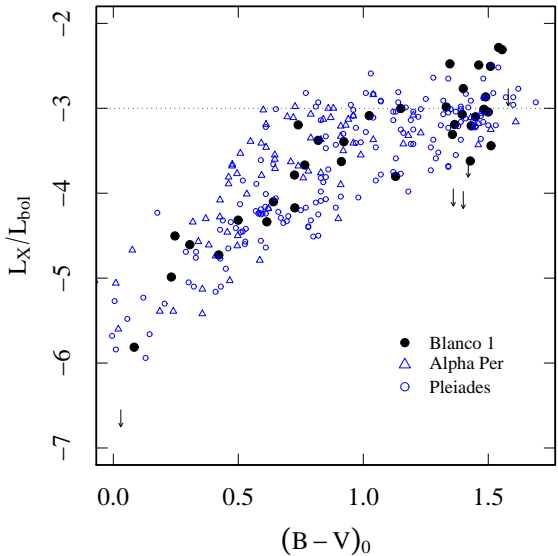}}
\caption{\label{lxlb} Ratio $\log
\mathrm{L}_\mathrm{X}/\mathrm{L}_\mathrm{bol}$
versus (B--V)$_0$ for Blanco\,1 cluster stars.
ROSAT data (detections only) of other clusters are also shown with
triangles, for $\alpha$\,Per \citep{Randich96}, and with circles, for
the Pleiades
\citep{Giusi999,Stauffer94}.
These data have been retrieved by usage of WEBDA 
database: http://obswww.unige.ch/webda}
\end{figure}

It is known that young, active stars exhibit a saturation of the ratio
$\log \mathrm{L}_\mathrm{X}/\mathrm{L}_\mathrm{bol}$ around the value -3
but it is not yet clear whether the saturation is due to the filling of
stellar surface 
with active coronal regions or to an intrinsic self-limiting mechanism
in the dynamo. 
The dependence of X-ray activity on age is thought to be due to the
relation 
between age and angular momentum losses: younger stars rotate more
rapidly and
as a consequence have more active X-ray coronae (\citealp{Nicola2003}
and references therein). 
\citet{Jardine2004} and \citet{Jardine99}, 
through modelling of coronae at high rotation rates, 
have suggested that centrifugal force 
can limit the size of coronal loops in fast rotators, thus explaining
the saturation phenomenon.
Studies of open clusters and star formation regions have shown that the
onset of
saturation in stars with ages of a few hundred million to a billion
years is observed as follows:
for stars with (B--V)$_0 \sim$ 0.6-0.7, saturation sets in at ages of a
few Myr up to 100 Myr (IC2391/IC2602, 
$\alpha$\,Per and Pleiades); for (B--V)$_0 \sim 1$, at 300 Myr (NGC
6475); and for (B--V)$_0 
\sim 1.4$, at 600 Myr (Hyades, Praesepe) \citep{Randich97}.
This pattern agrees with the rotation slow-down -- X-ray activity
decrease for
stars of greater and greater ages.
For Blanco\,1, we have calculated stellar 
$\log \mathrm{L}_\mathrm{X}/\mathrm{L}_\mathrm{bol}$ ratios with
L$_{bol}$ 
values by interpolating on (B--V)$_0$ from a 
100 Myr, $Z=0.03$ isochrone model \citep{Siess2000} (see Table
\ref{xmemb})\footnote{For one star, {\em P3}, no ratio is 
given because extrapolation to such a large B--V value was considered
unreliable.}.
As seen in Fig. \ref{lxlb}, the large scatter in Blanco\,1 dM stars with 
$\log \mathrm{L}_\mathrm{X}/ \mathrm{L}_\mathrm{bol}\geq -2.5$ obscures
any evidence of saturation in these
late type stars. 
However, nearly all of these objects underwent a flare-like event
during our observation, causing them to appear at higher $\log
\mathrm{L}_\mathrm{X}/\mathrm{L}_\mathrm{bol}$ values than
the remaining sample. Without these objects the distribution of
 $\log \mathrm{L}_\mathrm{X}/\mathrm{L}_\mathrm{bol}$ appears to flatten
around (B-V)$_0$ =1, equivalent to 0.8M$\sun$. Interpreted in this
manner, the $\mathrm{L}_\mathrm{X}/\mathrm{L}_\mathrm{bol}$ ratios for
Blanco\,1 are consistent
with those of other coeval clusters.

\begin{figure*}
\centering{
\includegraphics[height=8cm,angle=-90]{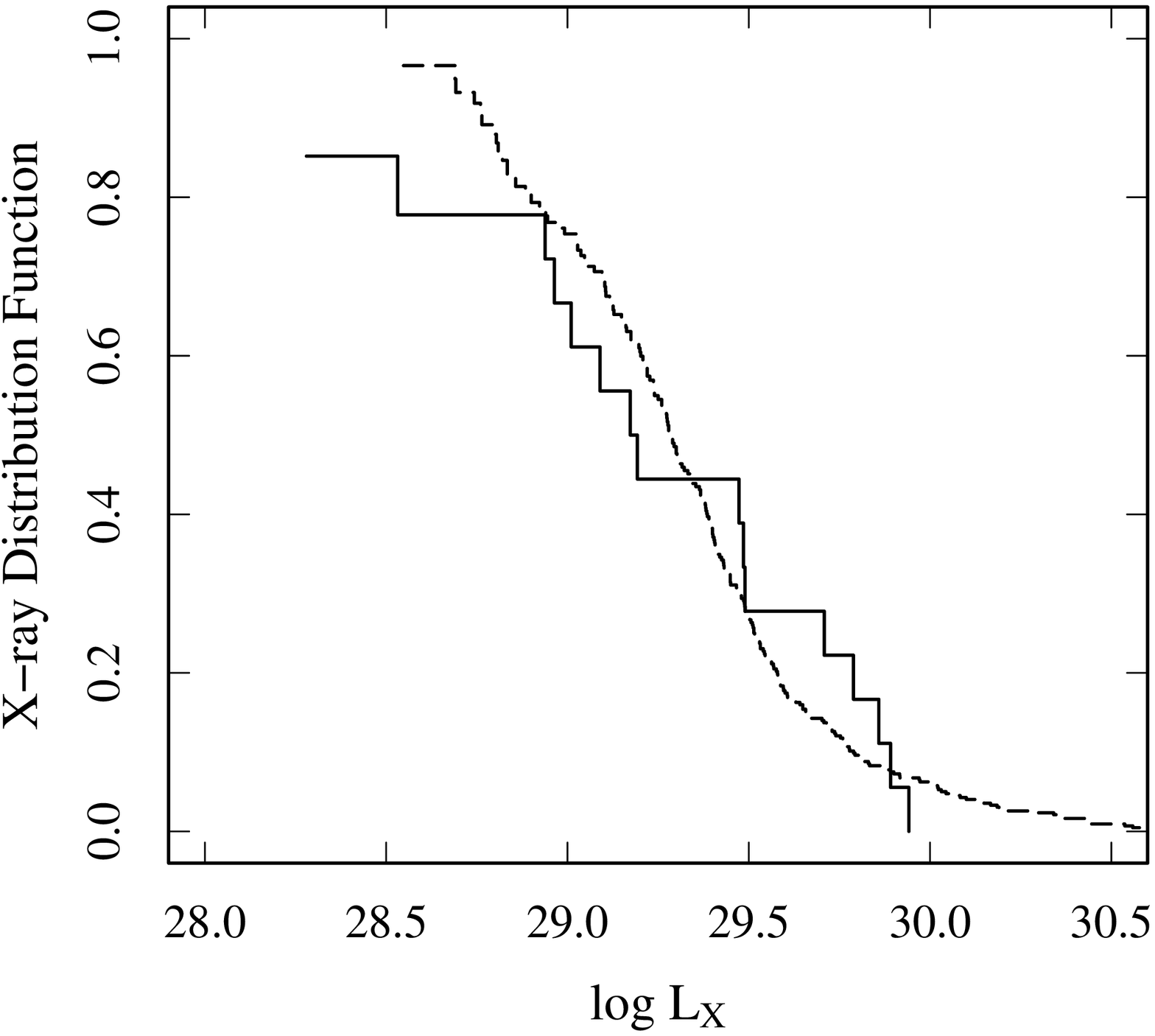}
\includegraphics[height=8cm,angle=-90]{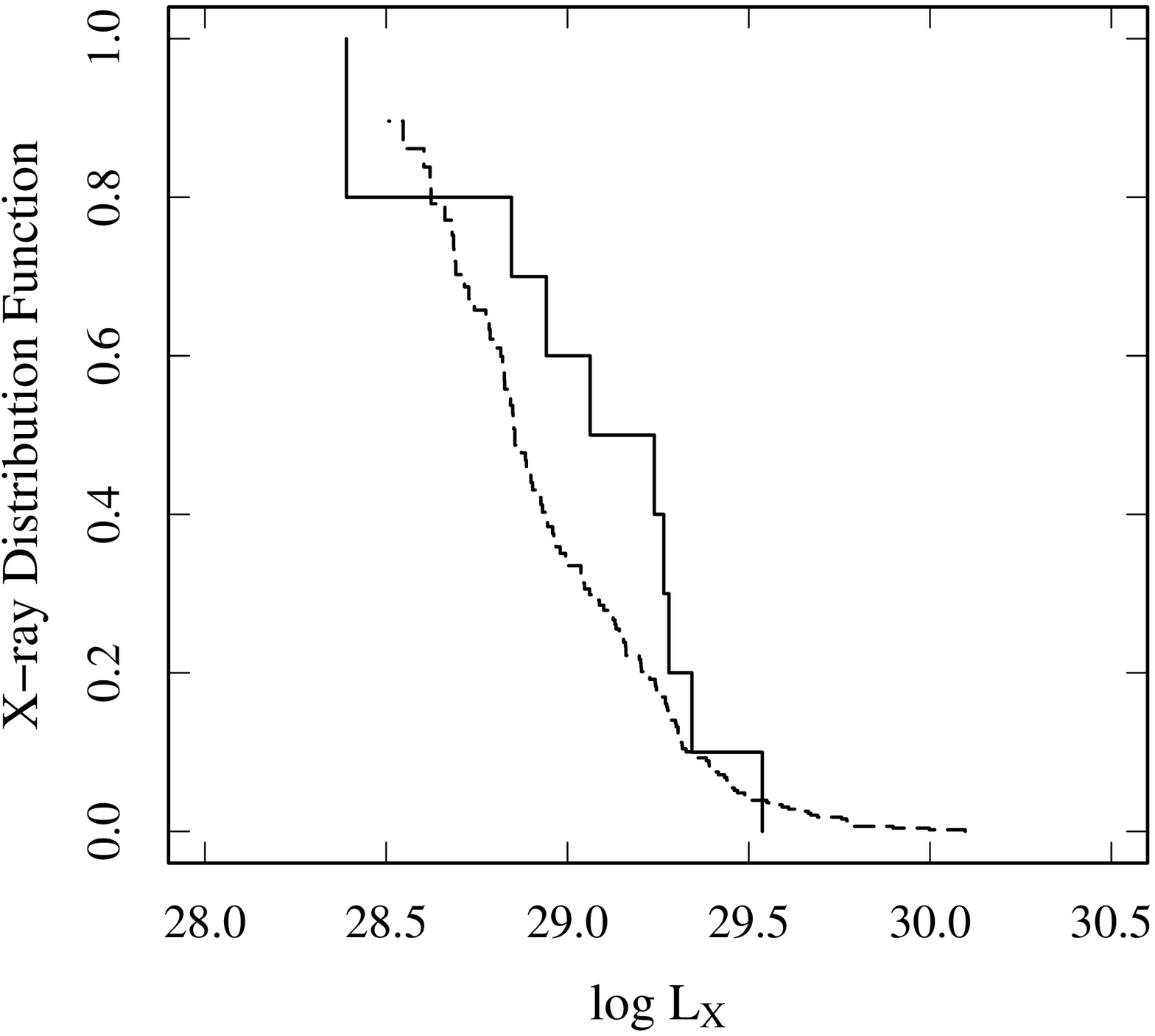}}
\caption{\label{xld} Kaplan-Meier estimator (solid line) of cumulative 
distribution functions (XLD) of $\log \mathrm{L_{X}}$
of dK (top panel) and dM stars (bottom panel). Dashed lines show
analogous ROSAT data for the Pleiades \citep{Giusi999,Giusi96},
comparison with these data are discussed in the text.}
\end{figure*}

The Kaplan-Meier estimators of $\log \mathrm{L_X}$ cumulative
distribution functions
for the censored samples of dK and dM stars are shown in Fig. \ref{xld}.
For comparison we report also the analogous Pleiades distributions 
from ROSAT data (dashed line).
These XLDs suggest a probable difference in the case of dM stars, as 
already reported in Paper I. 
In fact we observe that the XLD of dM stars 
in Pleiades and Blanco 1 overlap at high L$_X$ for values 
down to $\sim 29.4$, then the dM stars of Blanco 1 appears more
luminous, with a median of $\sim 29.2$ while Pleiades exhibit
a median of $\sim 28.8$.
A {\em Wilcoxon} two-sample test applied to the dM 
data of Blanco\,1 and the Pleiades show that the two distributions
differ at a confidence level of 97\%, while a {\em Gehan} test 
yields a difference at only the 94\% level
(see \citealp{Feigelson85} for a discussion of the Kaplan-Meier
estimator and two sample tests we used here).
If the difference is true, this could arise by the enhanced metallicity
of Blanco 1 stars with respect to the Pleiades ([Fe/H]=+0.23, \citealp{Edv95}).
The higher metallicity is expected to enhance the radiative losses, due to line
emission, which dominates the coronal spectra at 
temperatures of $\sim 0.3-1$ keV, such as those found in our spectral analysis.
  
The dK stars of Blanco\,1 have an emission level similar to that of the 
Pleiades, confirming the result already presented in Paper I.
As for dM stars, the dK XLD is well determined, 
with upper limits occurring only in the lower tail. 
For high-metal content dK stars \citep{NICOLA2001}, models predict
higher emission levels
than for
low metallicity dK stars, but we do not see this in our data.
Other factors, such as a different distribution of rotational
velocities,
might explain this apparent shortcoming of the models.

\section{Other X-ray sources and the completeness of cluster sequence}
\begin{figure}
\resizebox{\hsize}{!}{\includegraphics{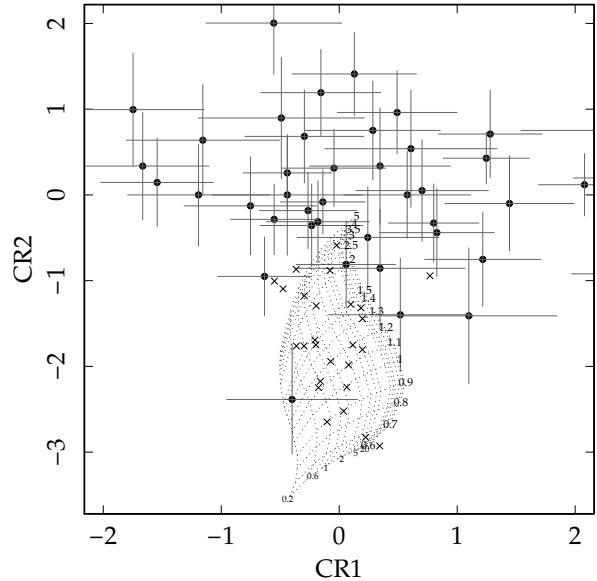}}
\caption{\label{unidcol} X-ray color diagram of the unidentified
sources.
For comparison, Blanco\,1 source data (crosses) and the model 
grid described in the Sect. \ref{color} are also shown.}
\end{figure}

The detection process revealed 64 X-ray sources with no known
counterpart in optical and infrared bands. 
In Table B.1 we give their positions, offsets, count rates and detection 
significance.
We have calculated for the brightest sources the X-ray colors
as described above. The color-color diagram in Fig. \ref{unidcol}
reveals that
most of these objects appear in a different region, well-separated from
cluster member sources, and implies they have different spectra.
Most of these sources are expected be extragalactic because
of the high latitude of this observation. 
Unidentified sources nr. 16 and 23 of Table B.1 do have X-ray colors
within the model grid region,
suggesting that their spectra are similar to those of cluster
coronae. These two objects may be unknown cluster stars with very low
mass but high X-ray activity.
The positions of sources nr. 39 and 51 fall within 10$\arcsec$
of two X-ray sources (BLX 35 and BLX 40) previously reported in Table 4
of \citet{Giusi99}.

\begin{figure*}
\centering{
\includegraphics[width=8cm, angle=-90]{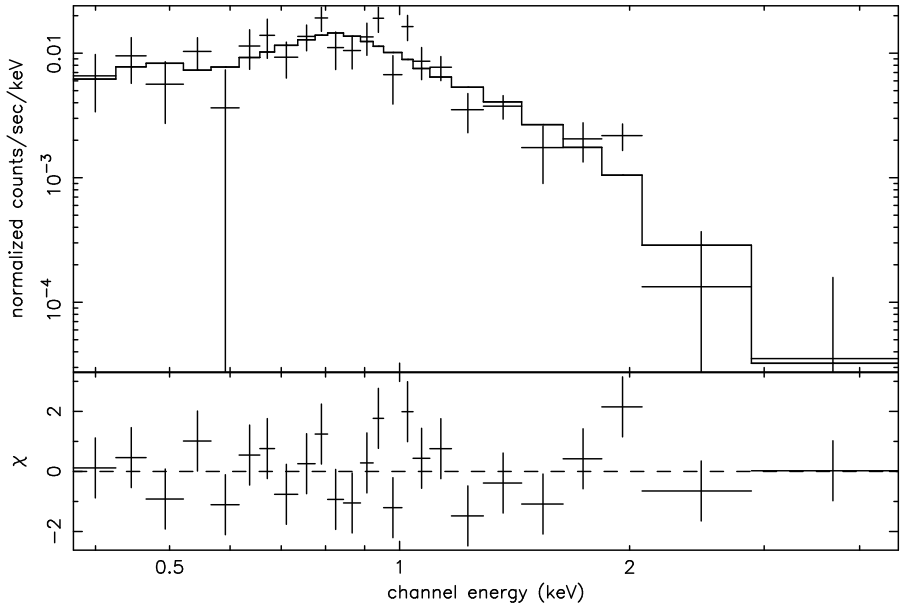}}
\caption{\label{x26fig} X-ray spectrum of the low mass flaring  
star nr. 26 in Table A.1}
\end{figure*}

We focused our attention on two of the sources with optical counterparts
not belonging to Blanco\,1:
the flaring X-ray source nr. 26 
and the star BLX 16 (respectively nr. 26 and 169 in Table A.1).  
In the proper motion analysis in Paper I, the optical 
counterpart of the source nr. 26 had a low probability of cluster
membership, 
and its photometry, consistent with low mass part of the cluster main
sequence,
suggests that this is a X-ray active, low mass field star.
The X-ray light curve exhibits a $\sim$20 ksec flare at the beginning of 
observation. %
Its spectral analysis (see Fig. \ref{x26fig})
favors an absorbed  1-T model with a
temperature around 0.65 keV and a N$_H$ value of $\sim 1.6\cdot 10^{21}$
cm$^{-2}$.
The abundance parameter is driven to very low value (0.03 $Z_{\sun} $) 
in order to obtain a good fit.
Its position in the X-ray color diagram (outside the region explored by
the 2-T models) also
suggests a different spectrum.

BLX 16 (nr. 169 in Table A.1) is an X-ray active source reported in 
\citet{Giusi99} with properties
similar to those of the lowest mass stars of the cluster.
In Paper I we discarded it as a cluster member because its position
in the color magnitude diagram was slightly off the cluster main
sequence.
We were able to assign a M1Ve spectral type from its spectrum
obtained in an optical follow-up, however, and with the present 
X-ray color data, we find its spectrum in agreement with that of the
other cluster 
stars. We therefore cannot rule out the possibility of this star as
another low mass member of the cluster. 
If accepted as a cluster member, its flux would be $\log f_\mathrm X =
-13.75$ 
erg sec$^{-1}$ cm$^{-2}$, and its X-ray luminosity would be $\log$ L$_X$
= 29.13, 
consistent with the L$_X$ XLD of dM stars in Fig. \ref{xld}.

\section{Summary}%
We have presented our analysis of an {\em XMM-Newton} observation
of the young open cluster Blanco\,1 addressing the
issues of the completeness of detection among X-ray cluster sources
and, for the first time, of low resolution X-ray spectroscopy of 
its members.

We have detected 190 X-ray sources by using a wavelet based algorithm
developed specifically for the EPIC camera.
Detections have been derived on the combined
MOS 1, 2 and {\em p--n} datasets, making full use of the EPIC camera
capability
and improving the detection sensitivity.
Of the 190 detected sources, 36 are identified with cluster members,
while 5 cluster members remain undetected.
As compared with previous ROSAT observations, the {\it XMM-Newton} rate
of detection has 
increased from 61\% to 88\%.

Six cluster sources with more than 1000 counts have been the subject of 
spectral analysis. A 2-T model fits the main 
spectrum features well and allows us to identify a ``cool'' component
around 
0.3 keV and a ``hot'' component around 1 keV, both with similar emission 
measures. These values are in good agreement with the XMM description of 
coronal spectra for young active stars like the Pleiades.
 
A 1-T model at $\sim 0.6$ keV explains the soft spectrum of an
intermediate
type star (spectral type $\sim$ A8) of the cluster. We consider that the
X-ray 
emission is intrinsic to the A star itself and not due to an unresolved
late
companion.

By means of two X-ray color indices, the 
X-ray spectral properties of a larger cluster sample were explored and
found to be consistent with the above 2-T model.

The X-ray luminosity distribution function of dM-type stars of the
cluster 
indicates a probable difference in emission levels with respect to the
Pleiades.
Saturation of $\log \mathrm{L}_\mathrm{X}/\mathrm{L}_\mathrm{bol}$
is not clearly evident down to mid-K stars, slightly later than in
Pleiades.

Appendix A lists non-cluster optical counterparts for 90 of the 190
X-ray sources, identified either with
the GSC-II and  USNO-B1 optical catalogs or with the 2MASS infrared
catalog.
One star is suggested as new probable low mass cluster member.

Listed in Appendix B are 64 sources that remain unidentified: 
most of them show X-ray colors
quite different from those of the cluster, suggesting a different
nature. However, for two of them the color analysis in Sect. 4 suggests
agreement of their spectrum  with those of cluster coronae.

\begin{acknowledgements}
IP, GM, SS and FD acknowledge financial support from ASI (Italian Space
Agency) and
MIUR (Ministero dell'Istruzione, dell'Universit\`a e della Ricerca); FRH
acknowledges
partial support from NASA grant NAG5-10005.\\
This research has made use of GSC-II, 2MASS, USNO-B1 catalogues through
the Vizier
database.
\end{acknowledgements}
\bibliographystyle{aa}
\bibliography{bibtesi}
\onecolumn
\appendix
\section{X-ray sources with optical counterparts in the GSC2, USNO-B1
and 2MASS catalogs.}
\begin{table}
\renewcommand{\arraystretch}{1.1}
\footnotesize
\caption{List of X sources with a counterpart in the optical and infrared catalogs 
USNO, GSC2 and 2MASS. The exposure times and the count rates are scaled on the MOS1 instrument.%
The final columns report the number of positional matches within 13$\arcsec$ in these catalogs.}
\begin{tabular}{l l l l l l l l l} \hline \hline
 Nr. Src X & RA  & Dec &  Pos. Err. & Exp. Time & Rate & Nr. opt. id. & Nr. opt. id. & Nr. opt. id. \\
 & (J2000) & (J2000) & arcsec & ksec. & cts/ksec. &  USNO & GSC2 & 2MASS \\ \hline
2 & 00 : 02 : 36.2 & -30 : 13 : 37.6 & 5.6 & 37.8 &$ 1.7 \pm 0.4 $& 3 & 1 & 0 \\
3 & 00 : 02 : 57.3 & -30 : 13 : 3.1 & 1.7 & 51.03 &$ 0.4 \pm 0.1 $& 1 & 0 & 0 \\
4 & 00 : 02 : 13.2 & -30 : 12 : 58.9 & 5.3 & 26.61 &$ 2.7 \pm 0.5 $& 1 & 0 & 0 \\
5 & 00 : 02 : 16.5 & -30 : 12 : 43.7 & 4.5 & 28.21 &$ 1.2 \pm 0.3 $& 1 & 0 & 0 \\
9 & 00 : 02 : 58 & -30 : 11 : 12.9 & 3.5 & 61.64 &$ 4.3 \pm 0.4 $& 1 & 0 & 0 \\
10 & 00 : 03 : 4.7 & -30 : 10 : 57.5 & 3.8 & 61.89 &$ 1 \pm 0.2 $& 8 & 1 & 0 \\
21 & 00 : 03 : 20.7 & -30 : 08 : 42.5 & 5.5 & 65.21 &$ 1 \pm 0.2 $& 11 & 1 & 1 \\
22 & 00 : 01 : 58.5 & -30 : 08 : 36 & 3.3 & 45.96 &$ 7.8 \pm 0.6 $& 2 & 0 & 0 \\
25 & 00 : 03 : 35.1 & -30 : 08 : 5.5 & 2.4 & 43.52 &$ 6.1 \pm 0.5 $& 1 & 0 & 0 \\
26 & 00 : 02 : 39.2 & -30 : 08 : 4.7 & 1.7 & 82.71 &$ 11.1 \pm 0.5 $& 1 & 1 & 1 \\
27 & 00 : 02 : 56.8 & -30 : 07 : 55.6 & 5.5 & 86.01 &$ 0.8 \pm 0.2 $& 2 & 0 & 0 \\
28 & 00 : 02 : 33.6 & -30 : 07 : 53.5 & 5.1 & 80.64 &$ 1 \pm 0.2 $& 4 & 1 & 1 \\
29 & 00 : 03 : 28.3 & -30 : 07 : 44.7 & 3.4 & 62.66 &$ 1.4 \pm 0.2 $& 1 & 0 & 1 \\
31 & 00 : 01 : 56.8 & -30 : 07 : 35.1 & 2.2 & 47.28 &$ 8 \pm 0.6 $& 1 & 1 & 0 \\
32 & 00 : 02 : 38 & -30 : 07 : 26.7 & 3.5 & 87.42 &$ 1 \pm 0.2 $& 1 & 0 & 0 \\
34 & 00 : 02 : 11.9 & -30 : 07 : 4.2 & 2.3 & 65.41 &$ 4.3 \pm 0.4 $& 1 & 0 & 0 \\
38 & 00 : 02 : 21.4 & -30 : 06 : 48 & 3 & 72.7 &$ 1.8 \pm 0.2 $& 1 & 0 & 0 \\
40 & 00 : 02 : 32 & -30 : 06 : 38.5 & 2.5 & 86.98 &$ 2.5 \pm 0.2 $& 1 & 0 & 0 \\
42 & 00 : 03 : 14.4 & -30 : 06 : 11.6 & 4.3 & 82 &$ 1.6 \pm 0.2 $& 1 & 0 & 0 \\
44 & 00 : 03 : 32.6 & -30 : 05 : 19.8 & 4.1 & 66.68 &$ 0.7 \pm 0.2 $& 2 & 0 & 0 \\
45 & 00 : 03 : 20.7 & -30 : 04 : 50.4 & 2.8 & 85.39 &$ 1.7 \pm 0.2 $& 1 & 1 & 1 \\
48 & 00 : 02 : 07.6 & -30 : 04 : 44.4 & 3.5 & 58.64 &$ 4.5 \pm 0.4 $& 2 & 2 & 2 \\
50 & 00 : 03 : 12.9 & -30 : 04 : 40.3 & 2.4 & 97.1 &$ 0.7 \pm 0.1 $& 1 & 0 & 1 \\
51 & 00 : 03 : 29.8 & -30 : 04 : 34.8 & 4.8 & 74.11 &$ 1.4 \pm 0.2 $& 1 & 1 & 1 \\
52 & 00 : 02 : 24.8 & -30 : 04 : 24.1 & 3.4 & 64.25 &$ 1.2 \pm 0.2 $& 1 & 1 & 1 \\
53 & 00 : 01 : 48.6 & -30 : 04 : 18.8 & 3.4 & 48.17 &$ 1.8 \pm 0.3 $& 1 & 0 & 0 \\
54 & 00 : 02 : 59.9 & -30 : 04 : 7.6 & 2.6 & 116.78 &$ 0.6 \pm 0.1 $& 2 & 0 & 0 \\
56 & 00 : 02 : 56.9 & -30 : 03 : 43.6 & 2.2 & 122.17 &$ 0.9 \pm 0.1 $& 1 & 1 & 0 \\
57 & 00 : 03 : 13.9 & -30 : 03 : 40.2 & 3.6 & 96.08 &$ 2.3 \pm 0.2 $& 1 & 1 & 1 \\
59 & 00 : 01 : 58.5 & -30 : 03 : 39.1 & 3.2 & 61.6 &$ 1.6 \pm 0.2 $& 3 & 0 & 0 \\
61 & 00 : 02 : 22.8 & -30 : 03 : 23.7 & 3.6 & 61.56 &$ 4.7 \pm 0.5 $& 1 & 0 & 0 \\
66 & 00 : 02 : 22.7 & -30 : 02 : 52.3 & 2.2 & 101.53 &$ 3.6 \pm 0.3 $& 1 & 1 & 1 \\
67 & 00 : 02 : 12.2 & -30 : 02 : 32.2 & 2.3 & 85.63 &$ 3.3 \pm 0.3 $& 1 & 0 & 0 \\
70 & 00 : 03 : 39.9 & -30 : 02 : 13 & 5.6 & 65.82 &$ 1 \pm 0.2 $& 2 & 1 & 1 \\
75 & 00 : 02 : 35 & -30 : 01 : 39.4 & 5.7 & 127.78 &$ 0.5 \pm 0.1 $& 3 & 1 & 1 \\
76 & 00 : 02 : 56.7 & -30 : 01 : 19.4 & 2.7 & 135.95 &$ 1.2 \pm 0.1 $& 3 & 1 & 0 \\
77 & 00 : 03 : 29.7 & -30 : 01 : 18.9 & 3.1 & 82.77 &$ 4.7 \pm 0.3 $& 3 & 2 & 2 \\
79 & 00 : 03 : 29.9 & -30 : 01 : 9 & 1.9 & 83.02 &$ 0.3 \pm 0.1 $& 2 & 2 & 2 \\
80 & 00 : 03 : 24.5 & -30 : 01 : 2.3 & 3.1 & 85.73 &$ 0.4 \pm 0.1 $& 5 & 1 & 2 \\
83 & 00 : 02 : 9.4 & -30 : 00 : 35.9 & 2.3 & 83.63 &$ 3.7 \pm 0.3 $& 2 & 0 & 1 \\
85 & 00 : 02 : 50.9 & -30 : 00 : 20.6 & 3.6 & 136.54 &$ 0.5 \pm 0.1 $& 2 & 0 & 0 \\
87 & 00 : 01 : 55.5 & -30 : 00 : 11.2 & 4.1 & 61.7 &$ 0.7 \pm 0.2 $& 1 & 0 & 0 \\
88 & 00 : 01 : 53.9 & -29 : 59 : 49.1 & 4.6 & 54.03 &$ 2 \pm 0.3 $& 2 & 1 & 1 \\
92 & 00 : 02 : 11.4 & -29 : 59 : 33 & 3 & 87.55 &$ 1.5 \pm 0.2 $& 1 & 0 & 0 \\
94 & 00 : 02 : 25.4 & -29 : 59 : 22.8 & 5.5 & 114.85 &$ 0.6 \pm 0.1 $& 3 & 2 & 2 \\
97 & 00 : 02 : 41.7 & -29 : 58 : 55.8 & 3.9 & 138.75 &$ 1.3 \pm 0.1 $& 1 & 1 & 1 \\
98 & 00 : 02 : 45.3 & -29 : 58 : 50.7 & 3.7 & 141.24 &$ 1.4 \pm 0.1 $& 6 & 1 & 1 \\
99 & 00 : 01 : 35.6 & -29 : 58 : 48.6 & 5 & 25.82 &$ 3.1 \pm 0.6 $& 1 & 1 & 1 \\
101 & 00 : 03 : 18.2 & -29 : 58 : 47.2 & 1.8 & 105.95 &$ 2.1 \pm 0.2 $& 1 & 1 & 0 \\
112 & 00 : 02 : 39 & -29 : 57 : 43.2 & 5.6 & 131.12 &$ 0.5 \pm 0.1 $& 3 & 0 & 0 \\
115 & 00 : 03 : 4.6 & -29 : 57 : 9.6 & 2.2 & 121.31 &$ 0.9 \pm 0.1 $& 1 & 0 & 0 \\
116 & 00 : 01 : 54.4 & -29 : 57 : 8.8 & 1.7 & 58.05 &$ 0.3 \pm 0.1 $& 3 & 1 & 1 \\
117 & 00 : 03 : 8.2 & -29 : 57 : 9.1 & 4.5 & 114.78 &$ 1 \pm 0.2 $& 1 & 0 & 0 \\ \hline
\end{tabular}
\end{table}
%
%
\begin{table}
\renewcommand{\arraystretch}{1.1}
\addtocounter{table}{-1}
\caption{Continued.}
\begin{tabular}{l l l l l l l l l} \hline \hline
 Nr. Src X & RA  & Dec &  Pos. Err. & Exp. Time & Rate & Nr. opt. id. & Nr. opt. id. & Nr. opt. id. \\
 & (J2000) & (J2000) & arcsec & ksec. & cts/ksec. &  USNO & GSC2 & 2MASS \\ \hline
119 & 00 : 02 : 10 & -29 : 56 : 48.2 & 3.3 & 55.43 &$ 0.6 \pm 0.2 $& 1 & 0 & 0 \\
120 & 00 : 03 : 1.8 & -29 : 56 : 46.1 & 2.9 & 122.38 &$ 0.3 \pm 0.1 $& 1 & 0 & 0 \\
127 & 00 : 02 : 35.1 & -29 : 55 : 57 & 1.7 & 116.24 &$ 7.7 \pm 0.3 $& 2 & 0 & 0 \\
130 & 00 : 02 : 55.9 & -29 : 55 : 37.8 & 4.1 & 120.09 &$ 1.4 \pm 0.2 $& 6 & 1 & 1 \\
131 & 00 : 03 : 11.1 & -29 : 55 : 20.7 & 3.1 & 87.74 &$ 1.2 \pm 0.2 $& 2 & 1 & 1 \\
135 & 00 : 02 : 41.7 & -29 : 55 : 10.5 & 4.1 & 115.99 &$ 0.4 \pm 0.1 $& 1 & 0 & 0 \\
136 & 00 : 02 : 45.2 & -29 : 55 : 3.6 & 2.2 & 117.02 &$ 3.2 \pm 0.2 $& 4 & 0 & 1 \\
138 & 00 : 02 : 30.8 & -29 : 54 : 48 & 3.9 & 103.77 &$ 0.5 \pm 0.1 $& 1 & 0 & 0 \\
139 & 00 : 01 : 49.1 & -29 : 54 : 47.2 & 4.1 & 47.41 &$ 3.5 \pm 0.4 $& 1 & 0 & 0 \\
141 & 00 : 02 : 44.3 & -29 : 54 : 41.8 & 2.8 & 112.57 &$ 0.4 \pm 0.1 $& 13 & 2 & 2 \\
143 & 00 : 03 : 20.9 & -29 : 54 : 35.7 & 4.1 & 84.59 &$ 0.6 \pm 0.1 $& 1 & 0 & 1 \\
144 & 00 : 02 : 8.6 & -29 : 54 : 32.5 & 3.5 & 69.55 &$ 1.1 \pm 0.2 $& 1 & 1 & 1 \\
145 & 00 : 03 : 5.2 & -29 : 54 : 24.8 & 1.5 & 87.17 &$ 0.3 \pm 0.1 $& 1 & 0 & 0 \\
146 & 00 : 03 : 2.6 & -29 : 54 : 23.7 & 1.8 & 105.89 &$ 7.1 \pm 0.3 $& 1 & 0 & 0 \\
147 & 00 : 02 : 36.2 & -29 : 54 : 15.7 & 4 & 94.62 &$ 0.5 \pm 0.1 $& 3 & 2 & 2 \\
148 & 00 : 01 : 54 & -29 : 54 : 5.6 & 2.9 & 51.82 &$ 2.7 \pm 0.3 $& 1 & 0 & 0 \\
150 & 00 : 03 : 22.6 & -29 : 53 : 52.2 & 2 & 78.48 &$ 6.9 \pm 0.4 $& 1 & 1 & 1 \\
151 & 00 : 02 : 38.4 & -29 : 53 : 48 & 2.5 & 103.99 &$ 2.2 \pm 0.2 $& 3 & 0 & 0 \\
153 & 00 : 02 : 0.9 & -29 : 53 : 47.4 & 4.6 & 58.74 &$ 1.8 \pm 0.3 $& 1 & 0 & 0 \\
154 & 00 : 02 : 18.8 & -29 : 53 : 28.4 & 5.7 & 79.43 &$ 0.8 \pm 0.2 $& 3 & 0 & 1 \\
155 & 00 : 02 : 41.7 & -29 : 53 : 0.9 & 3.6 & 97.51 &$ 0.7 \pm 0.1 $& 1 & 0 & 0 \\
156 & 00 : 03 : 36.6 & -29 : 52 : 57 & 4.4 & 26.05 &$ 1.1 \pm 0.3 $& 3 & 0 & 0 \\
158 & 00 : 02 : 2.1 & -29 : 52 : 38.6 & 2.9 & 49.35 &$ 2.9 \pm 0.4 $& 1 & 1 & 1 \\
161 & 00 : 03 : 5.6 & -29 : 52 : 0.4 & 3 & 82.74 &$ 1.6 \pm 0.2 $& 6 & 1 & 2 \\
165 & 00 : 03 : 9.3 & -29 : 51 : 38.7 & 3.6 & 77.89 &$ 0.9 \pm 0.2 $& 3 & 1 & 1 \\
166 & 00 : 02 : 59.4 & -29 : 51 : 32.9 & 3 & 71.74 &$ 0.6 \pm 0.1 $& 1 & 1 & 0 \\
169 & 00 : 02 : 23.4 & -29 : 50 : 40.4 & 2.6 & 68.55 &$ 3.1 \pm 0.3 $& 1 & 1 & 1 \\
170 & 00 : 02 : 18.4 & -29 : 50 : 32.4 & 4.2 & 61.95 &$ 0.7 \pm 0.2 $& 2 & 0 & 0 \\
172 & 00 : 02 : 26.7 & -29 : 50 : 14.6 & 3.3 & 68.27 &$ 1.3 \pm 0.2 $& 1 & 0 & 0 \\
174 & 00 : 03 : 0.7 & -29 : 49 : 44.5 & 3.3 & 61.98 &$ 5.5 \pm 0.5 $& 2 & 1 & 1 \\
178 & 00 : 02 : 52.6 & -29 : 49 : 36.9 & 5.2 & 72.69 &$ 1.1 \pm 0.2 $& 1 & 1 & 1 \\
180 & 00 : 02 : 21 & -29 : 49 : 19.6 & 3.7 & 41.42 &$ 5.8 \pm 0.6 $& 4 & 0 & 0 \\
182 & 00 : 02 : 14.6 & -29 : 49 : 3.5 & 3.2 & 20.61 &$ 13.8 \pm 1.1 $& 4 & 1 & 1 \\
185 & 00 : 03 : 3.8 & -29 : 48 : 19.9 & 2.5 & 61.46 &$ 1 \pm 0.2 $& 1 & 0 & 0 \\
186 & 00 : 02 : 35.9 & -29 : 48 : 14.4 & 3.6 & 60.39 &$ 4.1 \pm 0.4 $& 2 & 1 & 1 \\
188 & 00 : 02 : 43.3 & -29 : 47 : 39.5 & 3.5 & 48.31 &$ 0.6 \pm 0.2 $& 2 & 0 & 0 \\
190 & 00 : 02 : 47.9 & -29 : 46 : 36.9 & 4.6 & 23.82 &$ 4.1 \pm 0.6 $& 3 & 1 & 1 \\ \hline
\end{tabular}
\end{table}
\section{Unidentified X-ray sources}
\begin{table*}[h]
\renewcommand{\arraystretch}{1.1}
\caption{\label{unidtab} Table of X-ray unidentified X-ray source positions and properties.
The Name column follows the naming convention adopted for the unidentified {\em XMM} sources. The exposure times and the count rates are scaled on the MOS1 instrument.}
\renewcommand{\arraystretch}{1.1}
\begin{tabular}{l l l l l l l l} \hline \hline
 & Name & RA  & Dec & Offaxis & Significance & Time & Count rate \\ 
 &      & J2000 & J2000 & arcmin &  & ksec. & cts ksec$^{-1}$ \\ \hline
1 & XMMU J000142.4-295745 & 00 : 01 : 42.4 & -29 : 57 : 45.7 & 12.81 & 6.08 & 34.22 & 0.64  $\pm$  0.2 \\
2 & XMMU J000145-295441 & 00 : 01 : 45 & -29 : 54 : 41.5 & 12.97 & 25.69 & 28.67 & 13.85  $\pm$  1.08 \\
3 & XMMU J000146-295946 & 00 : 01 : 46 & -29 : 59 : 46.7 & 11.98 & 6.65 & 48.05 & 1.46  $\pm$  0.29 \\
4 & XMMU J000155.1-300200 & 00 : 01 : 55.1 & -30 : 02 : 9.2 & 10.44 & 27.82 & 59.54 & 7.33  $\pm$  0.5 \\
5 & XMMU J000158.2-295511 & 00 : 01 : 58.2 & -29 : 55 : 11.4 & 10.12 & 5.8 & 59.44 & 0.56  $\pm$  0.15 \\
6 & XMMU J000200.7-295118 & 00 : 02 : 00.7 & -29 : 51 : 18.3 & 11.8 & 17.2 & 41.55 & 34.7  $\pm$  5.2 \\
7 & XMMU J000200.1-300859 & 00 : 02 : 0.1 & -30 : 08 : 59.4 & 13.29 & 9.21 & 45.81 & 1.7  $\pm$  0.33 \\
8 & XMMU J000206.3-295820 & 00 : 02 : 6.3 & -29 : 58 : 20.4 & 7.61 & 13.2 & 77.49 & 1.52  $\pm$  0.2 \\
9 & XMMU J000210.8-295437 & 00 : 02 : 10.8 & -29 : 54 : 37.7 & 7.98 & 7.88 & 70.62 & 1.11  $\pm$  0.21 \\
10 & XMMU J000213.4-295600 & 00 : 02 : 13.4 & -29 : 56 : 3.6 & 6.77 & 10.08 & 76.78 & 1.22  $\pm$  0.2 \\
11 & XMMU J000214.2-295924 & 00 : 02 : 14.2 & -29 : 59 : 24.9 & 5.87 & 7.31 & 50.26 & 1  $\pm$  0.23 \\
12 & XMMU J000216.6-300300 & 00 : 02 : 16.6 & -30 : 03 : 8.8 & 6.68 & 7.48 & 90.64 & 0.66  $\pm$  0.13 \\
13 & XMMU J000219.3-295155 & 00 : 02 : 19.3 & -29 : 51 : 55.7 & 8.63 & 14.07 & 71.67 & 1.6  $\pm$  0.21 \\
14 & XMMU J000219.3-300015 & 00 : 02 : 19.3 & -30 : 00 : 15.7 & 4.89 & 8.64 & 103.13 & 0.7  $\pm$  0.12 \\
15 & XMMU J000219.5-295700 & 00 : 02 : 19.5 & -29 : 57 : 9.8 & 5.1 & 7.26 & 99.36 & 0.48  $\pm$  0.11 \\
16 & XMMU J000225.4-295614 & 00 : 02 : 25.4 & -29 : 56 : 14.5 & 4.49 & 22.81 & 102.27 & 2.43  $\pm$  0.2 \\
17 & XMMU J000226.8-300227 & 00 : 02 : 26.8 & -30 : 02 : 27.2 & 4.56 & 8.82 & 108.62 & 0.97  $\pm$  0.16 \\
18 & XMMU J000232.2-295941 & 00 : 02 : 32.2 & -29 : 59 : 41.2 & 2.04 & 7.91 & 127.29 & 0.72  $\pm$  0.13 \\
19 & XMMU J000232.5-300317 & 00 : 02 : 32.5 & -30 : 03 : 17.1 & 4.57 & 21.04 & 114.98 & 2  $\pm$  0.18 \\
20 & XMMU J000236.9-300916 & 00 : 02 : 36.9 & -30 : 9 : 16.9 & 10.2 & 4.97 & 70.53 & 0.76  $\pm$  0.19 \\
21 & XMMU J000237.3-295845 & 00 : 02 : 37.3 & -29 : 58 : 45 & 0.94 & 4.81 & 132.34 & 0.33  $\pm$  0.08 \\
22 & XMMU J000238.2-295624 & 00 : 02 : 38.2 & -29 : 56 : 24.1 & 2.81 & 5.51 & 122.69 & 0.17  $\pm$  0.05 \\
23 & XMMU J000243.8-300711 & 00 : 02 : 43.8 & -30 : 07 : 11.1 & 8.07 & 13.33 & 92.33 & 1.99  $\pm$  0.22 \\
24 & XMMU J000247-295347 & 00 : 02 : 47 & -29 : 53 : 47.9 & 5.47 & 5.67 & 105.8 & 0.29  $\pm$  0.08 \\
25 & XMMU J000251.2-300200 & 00 : 02 : 51.2 & -30 : 02 : 9.1 & 3.71 & 5.79 & 135.26 & 0.4  $\pm$  0.09 \\
26 & XMMU J000251.3-295544 & 00 : 02 : 51.3 & -29 : 55 : 44 & 4.04 & 11.13 & 115.06 & 0.89  $\pm$  0.13 \\
27 & XMMU J000252.8-295200 & 00 : 02 : 52.8 & -29 : 52 : 9.4 & 7.41 & 9.25 & 89.78 & 1.33  $\pm$  0.2 \\
28 & XMMU J000252.9-295700 & 00 : 02 : 52.9 & -29 : 57 : 7.4 & 3.22 & 5.44 & 132.34 & 0.26  $\pm$  0.07 \\
29 & XMMU J000253.7-295948 & 00 : 02 : 53.7 & -29 : 59 : 48.7 & 2.78 & 44.3 & 141.6 & 5.21  $\pm$  0.25 \\
30 & XMMU J000254-295800 & 00 : 02 : 54 & -29 : 58 : 8.8 & 2.92 & 10.3 & 136.83 & 0.74  $\pm$  0.11 \\
31 & XMMU J000254.2-300037 & 00 : 02 : 54.2 & -30 : 00 : 37.5 & 3.17 & 17.89 & 141.05 & 1.41  $\pm$  0.14 \\
32 & XMMU J000255-301200 & 00 : 02 : 55 & -30 : 12 : 5 & 13.29 & 5.48 & 56.69 & 0.78  $\pm$  0.2 \\
33 & XMMU J000255.4-301234 & 00 : 02 : 55.4 & -30 : 12 : 34.2 & 13.78 & 4.96 & 53.99 & 0.31  $\pm$  0.11 \\
34 & XMMU J000256.2-294943 & 00 : 02 : 56.2 & -29 : 49 : 43.7 & 9.94 & 5.9 & 63.3 & 0.58  $\pm$  0.14 \\
35 & XMMU J000256.8-301338 & 00 : 02 : 56.8 & -30 : 13 : 38.2 & 14.89 & 9.92 & 48.04 & 1.16  $\pm$  0.22 \\
36 & XMMU J000259.2-295514 & 00 : 02 : 59.2 & -29 : 55 : 14.3 & 5.5 & 7.55 & 115.44 & 0.42  $\pm$  0.08 \\
37 & XMMU J000300.2-295153 & 00 : 03 : 0.2 & -29 : 51 : 53.2 & 8.32 & 8.23 & 76.24 & 1.74  $\pm$  0.28 \\
38 & XMMU J000301.6-295031 & 00 : 03 : 1.6 & -29 : 50 : 31 & 9.68 & 4.86 & 68.44 & 0.5  $\pm$  0.14 \\
39 & XMMU J000301.8-295547 & 00 : 03 : 1.8 & -29 : 55 : 47.7 & 5.56 & 15.58 & 116.67 & 1.5  $\pm$  0.16 \\
40 & XMMU J000303.2-294900 & 00 : 03 : 3.2 & -29 : 49 : 5.8 & 11.11 & 7.91 & 65.83 & 1.25  $\pm$  0.24 \\
41 & XMMU J000304.1-300300 & 00 : 03 : 4.1 & -30 : 03 : 8.8 & 6.36 & 29.59 & 119.2 & 3.65  $\pm$  0.24 \\
42 & XMMU J000305.6-295910 & 00 : 03 : 5.6 & -29 : 59 : 10.4 & 5.27 & 9.16 & 62.55 & 0.71  $\pm$  0.16 \\
43 & XMMU J000306.3-294939 & 00 : 03 : 6.3 & -29 : 49 : 39.6 & 10.91 & 7.7 & 69.11 & 1.08  $\pm$  0.21 \\
44 & XMMU J000307.8-295800 & 00 : 03 : 7.8 & -29 : 58 : 3.2 & 5.85 & 10.96 & 120.61 & 0.94  $\pm$  0.13 \\
45 & XMMU J000308.2-300328 & 00 : 03 : 8.2 & -30 : 03 : 28 & 7.27 & 8.61 & 81.25 & 0.9  $\pm$  0.17 \\
46 & XMMU J000310.1-294837 & 00 : 03 : 10.1 & -29 : 48 : 37.3 & 12.22 & 7.67 & 60.47 & 0.81  $\pm$  0.17 \\
47 & XMMU J000311.6-300900 & 00 : 03 : 11.6 & -30 : 9 : 8.4 & 11.97 & 10.69 & 69.61 & 1.46  $\pm$  0.22 \\
48 & XMMU J000313.7-300047 & 00 : 03 : 13.7 & -30 : 00 : 47.8 & 7.23 & 15.26 & 114.37 & 1.33  $\pm$  0.16 \\
49 & XMMU J000314.3-294940 & 00 : 03 : 14.3 & -29 : 49 : 40.5 & 11.86 & 7.55 & 63.9 & 0.84  $\pm$  0.18 \\
50 & XMMU J000317.4-300643 & 00 : 03 : 17.4 & -30 : 06 : 43.5 & 10.91 & 10.03 & 78.87 & 1.42  $\pm$  0.21 \\ \hline
\end{tabular}
\end{table*}
\newpage
\begin{table*}[h]
\renewcommand{\arraystretch}{1.2}
\renewcommand{\textfraction}{0.2}
\addtocounter{table}{-1}
\caption{Continued.}
\begin{tabular}{l l l l l l l l} \hline \hline
 & Name & RA & Dec  & Offaxis & Significance & Time  & Count rate \\ 
 &      & J2000 & J2000 & arcmin &  & sec. & cts sec$^{-1}$  \\ \hline
51 & XMMU J000319.1-300219 & 00 : 03 : 19.1 & -30 : 02 : 19.1 & 8.78 & 27.49 & 98.67 & 3.6  $\pm$  0.26 \\
52 & XMMU J000319.4-294951 & 00 : 03 : 19.4 & -29 : 49 : 51.5 & 12.41 & 9.53 & 60.79 & 1.47  $\pm$  0.24 \\
53 & XMMU J000320.7-300149 & 00 : 03 : 20.7 & -30 : 01 : 49.1 & 8.96 & 16.56 & 97.08 & 2.01  $\pm$  0.21 \\
54 & XMMU J000321.1-300651 & 00 : 03 : 21.1 & -30 : 06 : 51.3 & 11.57 & 6.54 & 74.44 & 0.56  $\pm$  0.13 \\
55 & XMMU J000323.8-300340 & 00 : 03 : 23.8 & -30 : 03 : 40 & 10.26 & 5.64 & 84.17 & 0.74  $\pm$  0.17 \\
56 & XMMU J000326.2-300827 & 00 : 03 : 26.2 & -30 : 08 : 27.7 & 13.47 & 12.58 & 57.47 & 2.63  $\pm$  0.34 \\
57 & XMMU J000326.3-300449 & 00 : 03 : 26.3 & -30 : 04 : 49.9 & 11.29 & 25.72 & 77.5 & 4.39  $\pm$  0.33 \\
58 & XMMU J000330.5-295642 & 00 : 03 : 30.5 & -29 : 56 : 42.4 & 10.95 & 8.84 & 67.25 & 1.36  $\pm$  0.23 \\
59 & XMMU J000336.7-295200 & 00 : 03 : 36.7 & -29 : 52 : 3.8 & 13.94 & 5.32 & 23.17 & 1.07  $\pm$  0.31 \\
60 & XMMU J000337.4-300213 & 00 : 03 : 37.4 & -30 : 02 : 13.1 & 12.55 & 7.66 & 70.42 & 0.56  $\pm$  0.12 \\
61 & XMMU J000339.1-300842 & 00 : 03 : 39.1 & -30 : 08 : 42.4 & 15.76 & 6.6 & 34.4 & 1.81  $\pm$  0.41 \\
62 & XMMU J000340.2-295827 & 00 : 03 : 40.2 & -29 : 58 : 27.7 & 12.78 & 4.84 & 68.39 & 0.41  $\pm$  0.12 \\
63 & XMMU J000341.3-295725 & 00 : 03 : 41.3 & -29 : 57 : 25.6 & 13.11 & 14.14 & 27.3 & 4.26  $\pm$  0.54 \\
64 & XMMU J000344.2-300034 & 00 : 03 : 44.2 & -30 : 00 : 34.1 & 13.7 & 8.24 & 57.76 & 1.74  $\pm$  0.29 \\ \hline
\end{tabular}
\end{table*}
\newpage
\begin{figure*}
\includegraphics[height=25.2cm,width=18cm]{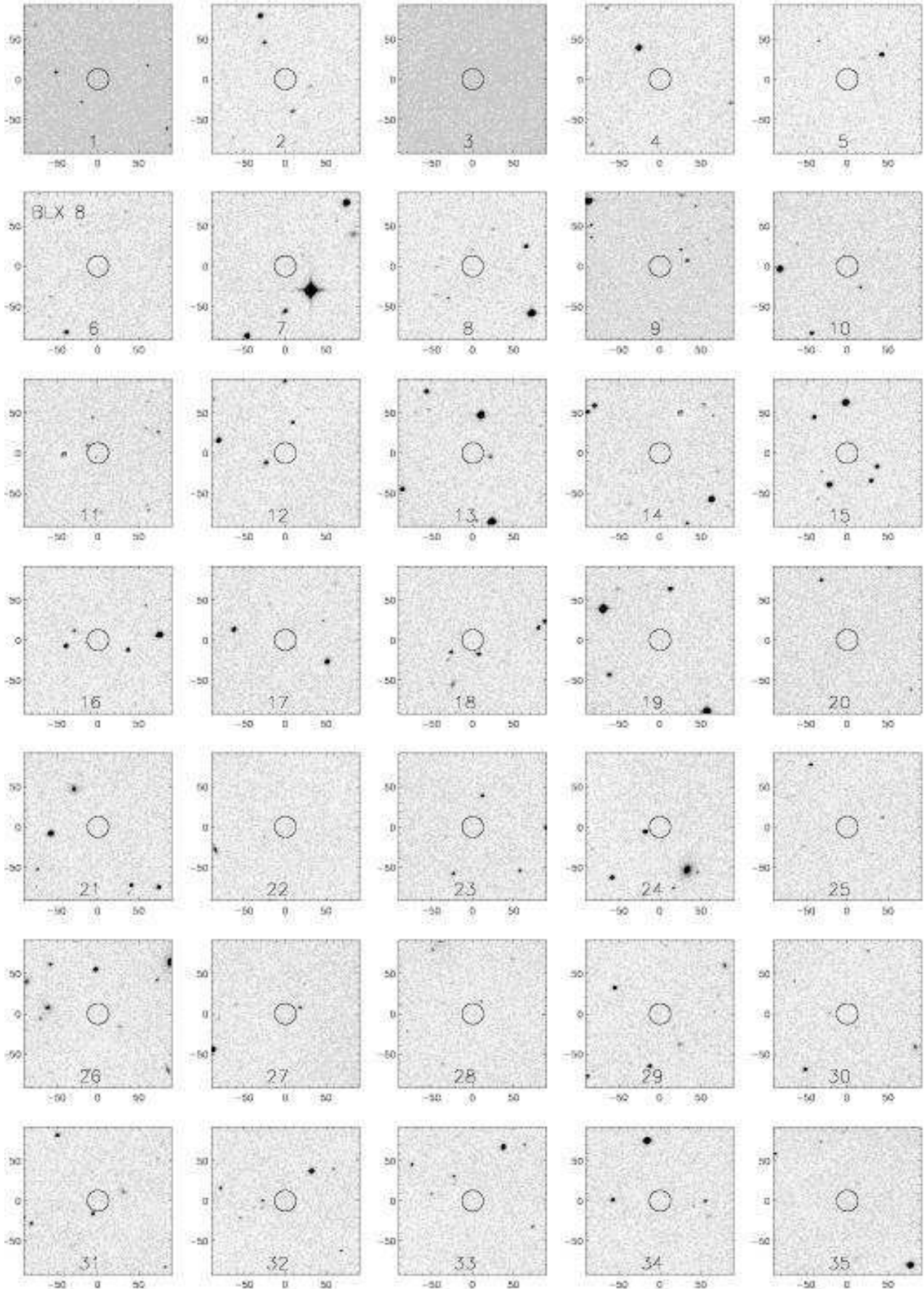}
\caption{\label{cha1} DSS2 -- IR finding charts (panel widths: 3' by 3') 
centered at X-ray positions of unidentified sources, with Table B.1
numbers indicated at bottom edge of each panel;
13$\arcsec$-radius circles indicate X-ray/optical matching criterion.}
\end{figure*}
\begin{figure*}
\includegraphics[height=25.2cm,width=18cm]{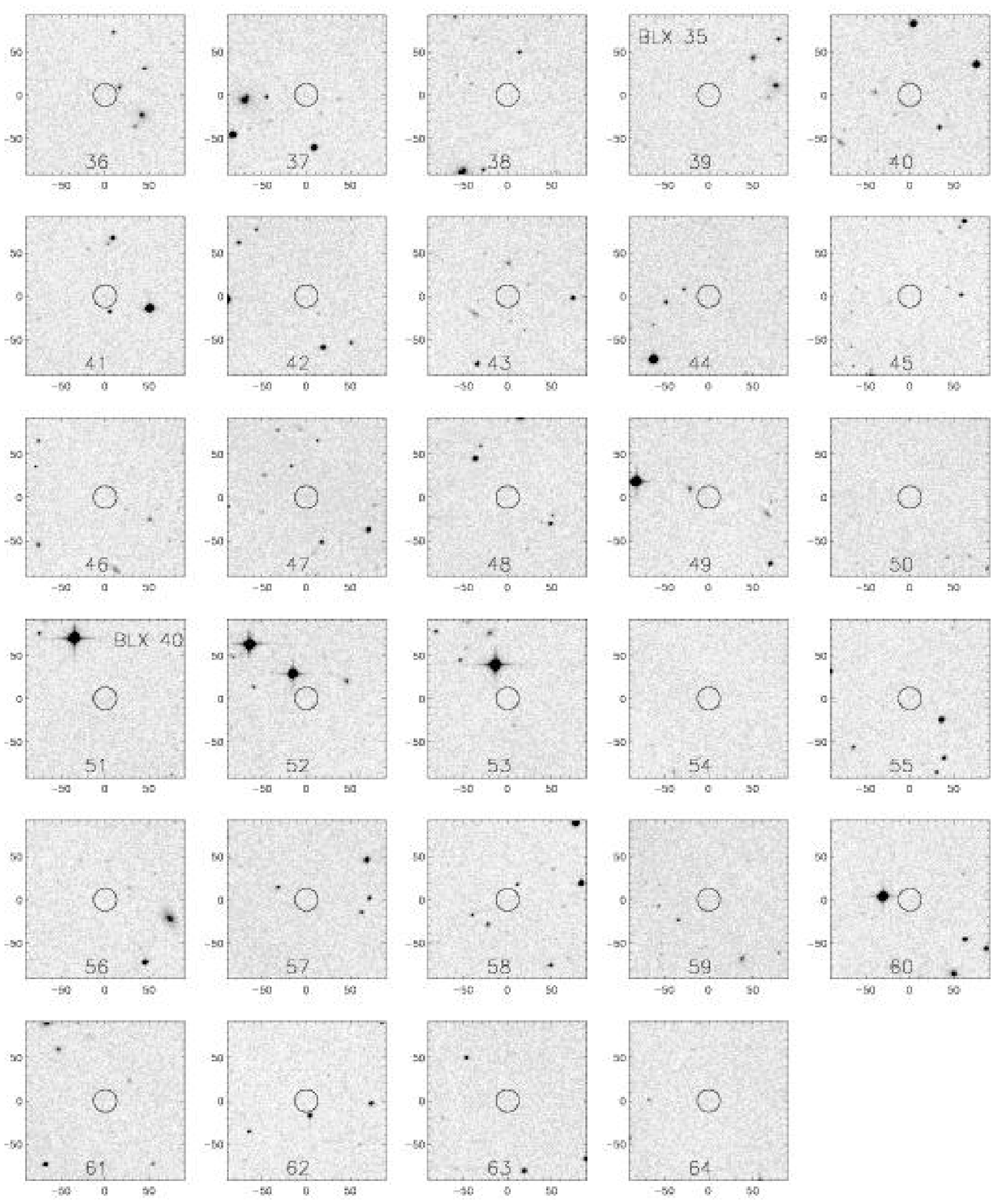}
\caption{As in Fig \ref{cha1}.}
\end{figure*}
\end{document}